\documentclass[preprintnumbers,amssymb, nofootinbib, nobibnotes,aps,prd]{revtex4}
\usepackage{soul, color}
\usepackage{amsmath}
\usepackage{braket}
\usepackage{graphicx}
\usepackage{dcolumn}
\usepackage{bm}
\usepackage{xcolor}

\newcommand{\beq}{\begin{equation}}
\newcommand{\eeq}{\end{equation}}
\newcommand{\bea}{\begin{eqnarray}}
\newcommand{\eea}{\end{eqnarray}}

\usepackage[unicode=true,pdfusetitle,
bookmarks=true,bookmarksnumbered=false,bookmarksopen=false,
breaklinks=false,pdfborder={0 0 1},backref=false,colorlinks=false]
{hyperref}

\begin{document}

\title{Strings, extended objects, and the classical double copy}
\author{Walter D. Goldberger}
\author{Jingping Li}
\affiliation{Physics Department, Yale University, New Haven, CT 06520, USA}
\date{\today}

\begin{abstract}

We extend Shen's recent formulation (arXiv:1806.07388) of the classical double copy, based on explicit color-kinematic duality, to the case of finite-size sources with non-zero spin.   For the case of spinning Yang-Mills sources, the most general consistent double copy consists of gravitating objects which carry pairs of spin degrees of freedom.   We find that the couplings of such objects to background fields match those of a classical (i.e. heavy) closed bosonic string, suggesting a string theory interpretation of sources related by color-kinematics duality.   As a special case, we identify a limit, corresponding to unoriented strings, in which the 2-form Kalb-Ramond axion field decouples from the gravitational side of the double copy.   Finally, we apply the classical double copy to extended objects, described by the addition of  finite-size operators to the worldline effective theory.   We find that consistency of the color-to-kinematics map requires that the Wilson coefficients of tidal operators obey certain relations, indicating that the extended gravitating objects generated by the double copy of Yang-Mills are not completely generic.
\end{abstract}
\maketitle

\section{Introduction}

About a decade ago, Bern, Carrasco, and Johansson (BCJ) conjectured~\cite{Bern:2008bcj1,Bern:2010bdhk} 
the existence of color-kinematics, or BCJ, duality of scattering
amplitudes in perturbative Yang-Mills theories.  Their work initiated the double
copy program that exploits this duality to construct perturbative amplitudes in
 gravity theories, generalizing the earlier KLT~\cite{Kawai:1986klt} relations between open and closed string scattering amplitudes.
 While the BCJ double copy has been rigorously proven for all tree-level amplitudes (see for instance \cite{proof}), the color-kinematic duality
remains conjectural for loop amplitudes \cite{Bern:2010bcj2}.  Nevertheless, many explicit checks of the conjecture at the loop level have been performed, leading to the recent study of five-loop amplitudes in Ref.~\cite{fiveloops}. For a comprehensive review of the current state of the field, see \cite{Bern:2019prr}.

In particular, the BCJ double copy leverages the relative simplicity of  gauge theory scattering amplitudes to improve
the efficiency of obtaining results in perturbative quantum gravity that would be otherwise intractable.  Naturally, one might also wonder if the double copy construction extends beyond amplitudes to other observables, for instance, to relate classical solutions of Yang-Mills and gravity theories.   This question was first raised by the work of~\cite{Monteiro:2014mow,Luna:2015lmow,Luna:2016hge} in the context of Kerr-Schild solutions to the Einstein field equations.   The Kerr-Schild double copy was further developed to more general geometries \cite{ksdc,ksdc2}, gravitational wave backgrounds \cite{wave}, and other applications \cite{etc}.   A review can be found in \cite{White:2017rev} as well as the later chapters of \cite{Bern:2019prr}.

A recent development in the classical double copy program is the application of these techniques to the case of compact binary inspirals in the perturbative regime. Such sources are well motivated by the recent  experimental breakthroughs in gravitational wave detection~\cite{ligovirgo}, where analysis of the data requires the construction of highly accurate templates.    Roughly, the determination of these gravitational wave templates can be divided into calculating a two-body Hamiltonian that describes the orbital dynamics, and the computation of gravitational radiation itself.   The computation of potentials from the double-copy of gauge theory amplitudes has been addressed in~\cite{CRS}, based on ideas introduced in~\cite{Neill:2013wsa,Vaidya:2014kza}, and extended to higher perturbative orders in~\cite{claspot}.   See also~\cite{scattering} for related work. 

The connection between classical radiating sources and the double copy was first studied in~\cite{Goldberger:2016ymdc}, which employed a worldline formalism (based on the effective field theory approach of~\cite{Goldberger:2006gr}, reviewed in~\cite{NRGRrev}) to describe the scattering of massive classical
point particles coupled respectively to Yang-Mills and gravitational fields.    By explicit calculation at leading order in perturbation theory, Ref.~\cite{Goldberger:2016ymdc} proposed a set of color-to-kinematic mapping rules that relate classical radiating solutions in gauge theory to those of a theory of gravity containing both graviton and scalar (dilaton) degrees of freedom.   The same mapping was found~\cite{Goldberger:2017bsdc} to generate Yang-Mills radiation from classical solutions of a certain bi-adjoint scalar~\cite{biadjoint} which plays a role in the color-kinematics duality of scattering amplitudes.   Spinning particles were studied in Refs.~\cite{Goldberger:2017cadc,Li:2018sgdc} where it was found necessary to include a two-form Kalb-Ramond axion gauge field $B_{\mu\nu}$ on the gravity side, with couplings to the graviton and dilaton that match those of low energy ``string gravity,'' in order to generate a consistent double copy.   Other recent developments in the application of the double copy to perturbative classical radiation include Refs.~\cite{Goldberger:2017bs,Plefka:2018psw} on bound states, Ref.~\cite{Luna:2017lnow} on connections to amplitudes and a proposal to remove dilaton contamination, and Ref.~\cite{Athira:2019} on connections to soft limits of scattering amplitudes.     In the context of radiation, a complementary line of attack that aims to relate the $S$-matrix directly to classical observables includes the recent work of Refs.~\cite{clasobs,scatangle,spin1,Bautista:2019sca,spin2,Bautista:2019evw,Kalin:2019rwq}.

Although the original proposal  in~\cite{Goldberger:2016ymdc} was successful in relating leading order solutions in various theories, the replacement rules presented there still left the connection to the original BCJ duality still somewhat unclear.   For instance, it was not known (as the authors of~\cite{Goldberger:2016ymdc} recognized) how the mapping rules could be applied to the more complicated color structures that would arise beyond leading order in the bi-adjoint and gauge theory solutions.   The situation was later clarified in a paper by Shen~\cite{Shen:2018nlo} which took as its starting point the setup of refs.~\cite{Goldberger:2016ymdc,Goldberger:2017bsdc} and used it to explicitly construct radiating solutions in bi-adjoint scalar theory, Yang-Mills, and gravity at the next order in perturbation theory.    The results revealed a factorization property of the radiation amplitudes which is not apparent at leading order, and which allows the application of color-kinematics duality to relate observables in different theories in a way that is much more closely related to the original BCJ duality of scattering amplitudes.    The approach proposed in~\cite{Shen:2018nlo}  provides an (in principle) all orders prescription for constructing double copies of classical systems, offers a more robust framework than the original proposal of~\cite{Goldberger:2016ymdc}, and has the potential to shed light on the microscopic principle underlying the double copy in the classical case.

Our goal in this paper is both to test and to generalize the more recent formulation of the classical double copy as presented in~\cite{Shen:2018nlo} away from the strict point particle limit.    To do this we consider first the case of spinning sources, working to linear order in the spin.   We find that when written in the factorized form advocated in~\cite{Shen:2018nlo}, the most general consistent double copy of the Yang-Mills radiation amplitude for spinning sources is a classical field sourced by objects carrying a pair of spin variables $S^{\mu\nu}_L$ and $S^{\mu\nu}_R$.   When the spins $S^{\mu\nu}_L$,  $S^{\mu\nu}_R$ are identified with one another, the axion field $B_{\mu\nu}$ decouples from the gravitational theory, at least to leading order in perturbation theory.   Remarkably, the case where the spin variables are treated as independent also has a consistent physical interpretation, as the low energy limit of a theory containing closed bosonic strings as sources. To verify this claim, we perform a multipole expansion of the term in the string worldsheet theory that couples to $B_{\mu\nu}$, which at leading order in derivatives is of the form
\begin{equation}
\int d^2 \sigma \epsilon^{ab} \partial_a X^\mu \partial_b X^\nu B_{\mu\nu}(X),
\end{equation}
finding precise agreement with the double copy prediction.    We also examine the proposal of~\cite{Shen:2018nlo} in the context of extended sources, which in the point particle limit which we study here, correspond to worldline Lagrangians containing higher dimensional operators that encode the response (color or tidal polarizability) of the source to applied long-wavelength fields.   In this case, we find that finite-size operators are consistent with the formulation of~\cite{Shen:2018nlo}, although there are constraints on the Wilson coefficients which reduce the number of independent operators, indicating that the double copy does not hold for generic extended objects.

We organize the paper as follows: in the next section, we briefly review the setup of classical charged worldlines coupled to scalar bi-adjoint, gauge, and gravitational fields introduced in refs.~\cite{Goldberger:2016ymdc,Goldberger:2017bsdc}, as well as the formulation of the classical double copy found in~\cite{Shen:2018nlo}.   In Sec. \ref{sec:spdc}, we examine the color-kinematic duality of radiation amplitudes from spinning sources, and show how the most general case can be interpreted as a ``duality'' of open strings acting as sources on the gauge theory side with closed strings coupled to gravity.    The case of higher dimensional operators encapsulating finite-size effects is summarized in~\ref{sec:hdim}, with computational details relegated to the appendix.   Finally, we present our conclusions in Sec. \ref{sec:disc}.

\section{The classical double copy}\label{sec:cdc}

We begin with a brief review of the double copy for perturbative classical systems, as proposed in Ref.~\cite{Goldberger:2016ymdc}.   For technical details, we refer the reader to the original literature~\cite{Goldberger:2016ymdc,Goldberger:2017bs}.

We consider a classical scattering problem where a set of point-like objects coupled to massless fields come in from infinity, interact via long range forces and emit radiation before escaping back to large separations.  The particular systems that we consider are the cubic bi-adjoint scalar field theory~\cite{biadjoint}, with an interaction term
\begin{equation}
\label{eq:bas}
S\supset-\frac{y}{3}\int d^{d}xf^{abc}\tilde{f}^{\tilde{a}\tilde{b}\tilde{c}}\phi^{a\tilde{a}}\phi^{b\tilde{b}}\phi^{c\tilde{c}},
\end{equation}
that is invariant invariant under $G\times {\tilde G}$ global symmetries acting on $\phi^{a{\tilde a}}$ in the bi-adjoint representation, Yang-Mills theory\footnote{Our sign conventions are $D_\mu = \partial_\mu + i g_s A^a_\mu T^a$,  $[T^a,T^b]=if^{abc} T^c$, $(T_{\mbox{\tiny{adj}}}^a)^b_c=-if^{abc}$.} with
\begin{equation}
S=-\frac{1}{4}\int d^{d}xF_{\mu\nu}^{a}F_{a}^{\mu\nu},
\end{equation}
and a gravity theory of fields $(\phi,B_{\mu\nu},g_{\mu\nu})$ whose action at the two-derivative level is defined by the ``string gravity'' Lagrangian in Einstein frame
\begin{equation}
S=-2m_{Pl}^{d-2}\int d^{d}x\sqrt{g}\bigg[R-(d-2)g^{\mu\nu}\partial_{\mu}\phi\partial_{\nu}\phi+\frac{1}{12} e^{-4\phi}H_{\mu\nu\rho}H^{\mu\nu\rho}\bigg],\label{eq:sgrav}
\end{equation}
with $H_{\mu\nu\rho} = (dB)_{\mu\nu\rho}$, which is known to play a role in the double copy of gauge theory amplitudes \cite{Bern:2016bdn} as well as in Kerr-Schild double copy~\cite{Luna:2016hge}.

The collection of classical sources are described by point particle effective field theories formulated on their worldlines.   Apart from the spacetime trajectory $x_\alpha^\mu(\tau)$, we assume that in the bi-adjoint theory each particle carries along its worldline a pair of color charges $c_\alpha^a(\tau),$ ${\tilde c}_\alpha^a(\tau),$ respectively transforming in the adjoint representations of $G$ and ${\tilde G}$. Given these variables, the lowest dimension interaction that is consistent with the symmetries is given by (suppressing particle labels)
\begin{equation}
\label{eq:bapp}
S_{pp} \supset y\int d\tau (c\cdot \phi\cdot {\tilde c})(x(\tau)),
\end{equation}
where $\tau$ is a reparametrization invariant time coordinate along the particle, and we have introduced the shorthand notation $c\cdot \phi\cdot {\tilde c}= (c\cdot \phi)^{\tilde a} {\tilde c}^{\tilde a} = c^a (\phi\cdot {\tilde c})^{a} = c^a \phi^{a\tilde a} {\tilde c}^{\tilde a}$. 

\begin{figure}
	\begin{centering}
		\includegraphics[scale=0.5]{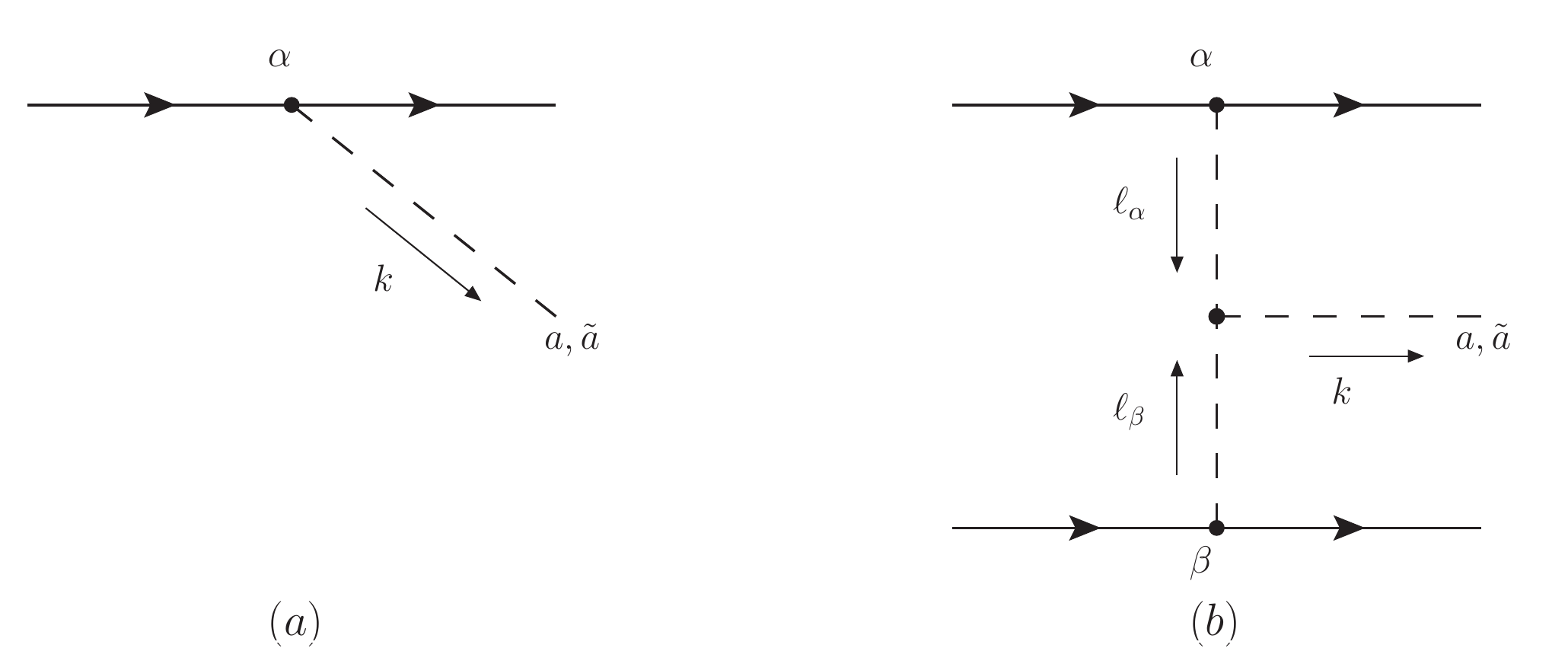}
		\par\end{centering}
	\caption{Calculation of massless bi-adjoint scalar radiation at leading order in perturbation theory. (a) corresponds to the worldline coupling with deflected trajectory via equations of motion, and (b) represents two worldline couplings and a cubic spacetime interaction. \label{fig:pert}}
\end{figure}

The scalar radiation emitted by a system of particles that interact via Eq.~(\ref{eq:bapp}) is computed in the perturbative regime. For this case, we explicitly show the corresponding Feynman diagrams in Fig.~\ref{fig:pert}. In the figure, we are depicting the perturbative corrections to the classical one-point function of $\phi^{a{\tilde a}}$ (the external line labeled by momentum $k^\mu$).   Vertex insertions correspond to either the cubic interaction in Eq.~(\ref{eq:bas}) or the worldline coupling in Eq.~(\ref{eq:bapp}).   In momentum space, the radiative part of this classical field is determined by an on-shell quantity which we denote ${\cal A}^{a\tilde a}(k)$, with $k^2=0$. It is defined as the tree-level probability amplitude to emit a single scalar particle of momentum $k$ by the classical point sources.    Equivalently, ${\cal A}^{a\tilde a}(k)$ is proportional to the coefficient of $|{\vec x}|^{1-d/2}$ in the classical solution $\phi^{a\tilde a}(\omega,{\vec x})=\int dx^0 e^{i\omega x^0} \phi^{a\tilde a}(x^0,{\vec x})$ as $|{\vec x}|\rightarrow \infty$, where the momentum is given by $k^\mu = \omega (1,{\vec x}/|{\vec x}|)$.

Ref.~\cite{Goldberger:2017bsdc} computed the classical solution for a system of particles interacting through Eq.~(\ref{eq:bapp}) in a configuration where they start out well separated at $\tau\rightarrow -\infty$, emit radiation at intermediate times, and scatter out to spatial infinity at $\tau\rightarrow \infty$.   In terms of the initial momenta (rescaled by the masses) $p^\mu_\alpha\equiv dx^\mu_\alpha/d\tau\vert_{\tau\rightarrow -\infty}$ and impact parameters $b^\mu_\alpha$, defined by the asymptotic orbits
\begin{equation}
\label{eq:init}
x_\alpha^\mu(\tau\rightarrow-\infty) = b^\mu_\alpha + p^\mu_\alpha \tau,
\end{equation} 
and the initial color charges $c_{\alpha}^{a},$ ${\tilde c}_{\alpha}^{\tilde a}$ at $\tau\rightarrow -\infty$, the leading order perturbative amplitude can be put in the form
\begin{align}
\mathcal{A}^{a\tilde{a}}(k) & =-iy^{3}\sum_{\alpha,\beta}\int_{{\alpha,\beta}}\mu_{\alpha,\beta}(k)\bigg[((c_{\alpha}\cdot c_{\beta}) c_{\alpha})^{a}\ell_{\alpha}^{2}\frac{k\cdot\ell_{\beta}}{(k\cdot p_{\alpha})^{2}}((\tilde{c}_{\alpha}\cdot\tilde{c}_{\beta})\tilde{c}_{\alpha})^{\tilde{a}}\nonumber \\
& +[c_{\alpha},c_{\beta}]^{a}\frac{\ell_{\alpha}^{2}}{k\cdot p_{\alpha}}((\tilde{c}_{\alpha}\cdot\tilde{c}_{\beta})\tilde{c}_{\alpha})^{\tilde{a}}+((c_{\alpha}\cdot c_{\beta})c_{\alpha})^{a}\frac{\ell_{\alpha}^{2}}{k\cdot p_{\alpha}}[\tilde{c}_{\alpha},\tilde{c}_{\beta}]^{\tilde{a}}+[c_{\alpha},c_{\beta}]^{a}(-1)[\tilde{c}_{\alpha},\tilde{c}_{\beta}]^{\tilde{a}}\bigg],\label{eq:bs}
\end{align}
where we have introduced the short hand notation for the integrals $\int_\alpha = \int d^d \ell_\alpha/(2\pi)^d$, and
\begin{equation}
\mu_{\alpha,\beta}(k) =\left[(2\pi)\delta(\ell_{\alpha}\cdot p_{\alpha})\frac{e^{i\ell_{\alpha}\cdot b_{\alpha}}}{\ell_{\alpha}^{2}}\right]\left[(2\pi)\delta(\ell_{\beta}\cdot p_{\beta})\frac{e^{i\ell_{\beta}\cdot b_{\beta}}}{\ell_{\beta}^{2}}\right](2\pi)^{d}\delta^{d}(\ell_{\alpha}+\ell_{\beta}-k).
\end{equation}  

Similarly, in Yang-Mills theory, Ref.~\cite{Goldberger:2016ymdc} analyzed the perturbative radiation field sourced by an ensemble of point color charges $(x^\mu(\tau),c^a(\tau))$, each minimally coupled to the the gauge field by a worldline term
\begin{equation}
\label{eq:ympp}
S_{pp}\supset -g \int dx^\mu (c\cdot A)^\mu(x(\tau)).
\end{equation}
In this case, the long-distance radiation emitted by the collection of interacting charges is  $\epsilon(k) \cdot {\cal A}^{a}(k),$ where the emission amplitude corresponding to the kinematics in Eq.~(\ref{eq:init}), and initial charges $c^a_\alpha$ is given by an expression analogous to Eq.~(\ref{eq:bs}),
\begin{align}
\mathcal{{A}}^{a\mu}(k)= & -ig^{3}\sum_{\alpha,\beta}\int_{\ell_{\alpha,\beta}}\mu_{\alpha,\beta}(k)\bigg[((c_{\alpha}\cdot c_{\beta})c_{\alpha})^{a} \ell_{\alpha}^{2}\frac{k\cdot\ell_{\beta}}{(k\cdot p_{\alpha})^{2}}(p_{\alpha}\cdot p_{\beta})p_{\alpha}^{\mu}+[c_{\alpha},c_{\beta}]^{a}\frac{\ell_{\alpha}^{2}}{k\cdot p_{\alpha}}(p_{\alpha}\cdot p_{\beta})p_{\alpha}^{\mu}\nonumber \\
& +(c_{\alpha}\cdot c_{\beta})c_{\alpha}^{a}\frac{\ell_{\alpha}^{2}}{k\cdot p_{\alpha}}\Big\{(k\cdot p_{\alpha})p_{\beta}^{\mu}-(k\cdot p_{\beta})p_{\alpha}^{\mu}-(p_{\alpha}\cdot p_{\beta})\ell_{\beta}^{\mu}\Big\}\nonumber \\
& +[c_{\alpha},c_{\beta}]^{a} (-1) \bigg\{(k\cdot p_{\alpha})p_{\beta}^{\mu}-(k\cdot p_{\beta})p_{\alpha}^{\mu}-{1\over 2}(p_{\alpha}\cdot p_{\beta})(\ell_{\beta}-\ell_\alpha)^{\mu}\bigg\}\bigg],\label{eq:ymns}
\end{align}
which is defined up to pure gauge terms proportional to $k^\mu$ that do not contribute for on-shell polarizations $k\cdot\epsilon(k)=0$. (The diagrams are similar to those of Fig.~\ref{fig:pert}, so we omit them here.)

Motivated by BCJ duality~\cite{Bern:2008bcj1}, and the work of~\cite{Monteiro:2014mow} on the non-perturbative double copy, Ref.~\cite{Goldberger:2016ymdc} considered the effect of the following set of formal substitution rules 
\begin{eqnarray}
\nonumber
c^a &\mapsto& p^\mu_\alpha\\
\label{eq:rules}
[c_\alpha,c_\beta]^a  & \mapsto & \Gamma_{\mu\nu\rho}(-k,\ell_{\alpha},\ell_{\beta}) p_\alpha^\nu p_\beta^\rho \equiv\frac{1}{2}(p_\alpha\cdot p_\beta) \left[(\ell_{\beta}-\ell_{\alpha})^{\mu}+(2 k\cdot p_\beta - p_\beta\cdot \ell_\beta) p_\alpha^\mu - (2 k\cdot p_\alpha-\ell_\alpha\cdot p_\alpha) p^\mu_\beta\right],\\
\nonumber
g & \mapsto & \kappa\equiv 1/(2m_{Pl})^{(d-2)/2},
\end{eqnarray}
on the Yang-Mills amplitude ${\cal A}^{a\mu}(k)$.   Under these color-to-kinematics transformations, one finds ${\cal A}^{a\mu}(k)\mapsto {\cal A}^{\mu\nu}(k)$ that can be contracted to two independent polarization tensors $\epsilon^\mu(k)\tilde{\epsilon}^\nu(k)$ to obtain an emission amplitude. We note that ${\cal A}^{a\mu}(k)$ is defined up to terms that vanish on-shell, so as shown in Ref.~\cite{Goldberger:2016ymdc}, this ambiguity was used to add an ``improvement term'' to arrive at the Yang-Mills amplitude in Eq.~(\ref{eq:ymns}) such that the predicted ${\cal A}^{\mu\nu}(k)$ obeys the Ward identities
\begin{equation}
k_\mu {\cal A}^{\mu\nu}(k) = k_\nu {\cal A}^{\mu\nu}(k) =0,
\end{equation}
for on-shell momenta.

 Because the amplitude ${\cal A}^{\mu\nu}(k)$ has at most double poles at $\ell_\alpha^2=\ell_\beta^2=0$, it can be interpreted as the radiation amplitude in a local theory of massless fields with cubic interactions.   Decomposing the product $\epsilon_\mu {\tilde \epsilon}_\nu$ into symmetric traceless tensor $\epsilon_{\mu\nu}(k)\equiv\epsilon_{(\mu} {\tilde \epsilon}_{\nu)}-\pi_{\mu\nu}$, anti-symmetric tensor $a_{\mu\nu}(k)\equiv \epsilon_{[\mu} {\tilde \epsilon}_{\nu]}$, and scalar $\pi_{\mu\nu}(k)\equiv\frac{\epsilon\cdot\tilde{\epsilon}}{d-2}\big(\eta_{\mu\nu}-\frac{k_\mu q_\nu+k_\nu q_\mu}{k\cdot q}\big)$ (where $q$ is an arbitrary reference null vector) little group representations, one finds by direct calculation that ${\cal A}^{\mu\nu}(k)$ describes radiation in a theory of spinless point particles coupled to graviton $g_{\mu\nu}$ and dilaton $\phi$ bulk modes with Lagrangian given in Eq.~(\ref{eq:sgrav}) (with $H_{\mu\nu\rho}=0$ due to the tensor symmetry ${\cal A}^{(\mu\nu)}=0$).  To leading order, one finds that the sources are described by the worldline action 
\begin{equation}
\label{eq:gpp}
S_{pp}\supset -m\int e^{\phi} \sqrt{g_{\mu\nu} dx^\mu dx^\nu}
\end{equation}
 and therefore couple naturally to the Weyl re-scaled string frame metric ${\tilde g}_{\mu\nu} = e^{2\phi} g_{\mu\nu}$.

 The results in~\cite{Goldberger:2016ymdc} were later extended~\cite{Goldberger:2017bsdc} to show that the same color-to-kinematic substitutions spelled out in Eq.~(\ref{eq:rules}) also generate the gauge theory amplitude $\epsilon^a(k)\cdot {\cal A}^{a}(k)$ from the bi-adjoint result ${\cal A}^{a\tilde a}(k)$ in Eq.~(\ref{eq:bs}).   Furthermore, as shown in~\cite{Goldberger:2017cadc,Li:2018sgdc}, the mapping rules are also consistent if particle spin degrees of freedom $S^{\mu\nu}$ with additional chromo-magnetic interactions
\begin{equation}
\label{eq:dipole}
S_{pp}\supset - {g\over m} \int d\tau S^{\mu\nu} (c\cdot F)_{\mu\nu}
\end{equation}
are included on the gauge theory side.  In this case (and only for the precise numerical value of the magnetic coupling in Eq.~(\ref{eq:dipole})) the double copy includes radiation in the axion channel as well as dilaton and graviton radiation.   The spin $S^{\mu\nu}$ (defined here to obey the constraint $S^{\mu\nu} p_\nu =0$) interacts with fields $(\phi,g_{\mu\nu},B_{\mu\nu})$ through a term of the form 
\begin{equation}
\label{eq:spin}
S_{pp}\supset {1\over 2} \int dx^\rho e^{-2\phi} \left[\Gamma_{\mu\nu\rho} + {1\over 2}  H_{\mu\nu\rho}\right]  S^{\mu\nu},
\end{equation}
at least to linear order in the spin variable.   Note that the magnitude of the axion coupling is a prediction of the double copy map in Eq.~(\ref{eq:rules}).   As in the spinless case, the interaction in Eq.~(\ref{eq:spin}) can be given a simpler geometric interpretation in terms of the ``string frame'' spin ${\tilde S}_{\mu\nu} = e^{2\phi} S_{\mu\nu}$, which couples as
\begin{equation}
S_{pp}\supset  {1\over 2} \int dx^\rho  {C^+}_{\mu\nu\rho} {\tilde S}^{\mu\nu}.\label{eq:pcp}
\end{equation}
Here the symmetric part ${C^+}^\rho_{(\mu\nu)} = {\tilde \Gamma}^\rho{}_{\mu\nu}$ corresponds to the Christoffel symbol of the string frame metric ${\tilde g}_{\mu\nu} = e^{2\phi} g_{\mu\nu}$ and the torsion $T^\rho{}_{\mu\nu} = {C^+}^\rho{}_{[\mu\nu]}= e^{-2\phi} H^\rho{}_{\mu\nu} \equiv  {\tilde H}^\rho{}_{\mu\nu}$ corresponds to the axion field strength. We will elaborate on the spinning results in Sec~\ref{sec:spdc}.

\subsection{Beyond leading order and color-kinematic duality}

Although the double copy procedure as summarized in the previous section yields well defined relations between leading order bi-adjoint, gauge, and gravitational radiation, color-kinematic duality in the original BCJ sense is not explicit for the gauge theory result. Due to this reason, it is (as already noted in Ref.~\cite{Goldberger:2016ymdc}) not clear from these results wether the rules summarized in Eq.~(\ref{eq:rules}) are sufficient to generate consistent predictions at higher orders in perturbation theory. For example, at the next to leading order (NLO), the Yang-Mills amplitude ${\cal A}^{\mu a}$ involves color structures of the form~\cite{Shen:2018nlo} 
\begin{equation}
\begin{array}{ccccc}
(c_\alpha\cdot c_\beta) (c_\alpha\cdot c_\gamma)c^a_\alpha,   & (c_\alpha\cdot c_\beta) (c_\alpha\cdot c_\gamma)c^a_\beta, &  (c_\alpha\cdot [c_\beta, c_\gamma]) c_\beta^a , & (c_\alpha\cdot c_\beta) [c_\alpha,c_\gamma]^a, &  [[c_\alpha, c_\beta],c_\gamma]^a , 
\end{array}
\end{equation}
and it is not obvious that Eq.~(\ref{eq:rules}) gives a well defined prescription for converting these objects into consistent kinematic structures.

Recently, Ref.~\cite{Shen:2018nlo} computed the NLO corrections to the radiation amplitudes in the theories defined in the previous section, coupled to spinless particles. Based on the explicit NLO results,~\cite{Shen:2018nlo} proposed a generalization of the classical double copy rules Eq.~(\ref{eq:rules}) which explicitly makes use of color-kinematic duality, naturally extending the classical double copy construction to higher orders in perturbation theory.

The proposal takes as its starting point the $n$-th order perturbative amplitude ${\cal A}_n^{a\tilde a}$ in the bi-adjoint theory, expressed as a sum over independent ``color numerators'' $C^a_i$, ${\tilde C}^{\tilde a}_i$, schematically\footnote{For notational clarity, we have absorbed the sum over particle labels into the indices $i,j$ that run over color structures.},
\begin{equation}
\label{eq:snlo}
{\cal A}^{a\tilde a}_n(k) = y^n \sum_{ij} C^a_i P^{ij}(k) {\tilde C}_j^{\tilde a}.
\end{equation}
This equation defines a non-diagonal  ``propagator matrix'' $P^{ij}(k),$ which depends on a mixture of poles from both the worldline and spacetime propagators.   This differs from the propagator factors appearing in the case of BCJ duality for scattering amplitudes, which in this context can be regarded as a matrix proportional to the identity.    A similar decomposition for the gauge theory amplitude at NLO was also found
\begin{equation}
\label{eq:yma}
{\cal A}^{\mu a}_n(k) = g^n \sum_{ij} C^a_i P^{ij}(k) N_j^\mu(k),
\end{equation}
where the ``kinematic numerators'' $N_i$ obey the Ward identity $k_\mu N^\mu_i(k) =0$. It is important to note that the $N_i^\mu(k)$ are chosen in a way such that they obey similar particle interchange symmetries and kinematic Jacobi identities as the corresponding color objects $C^a_i$ do.   In this sense the map between bi-adjoint and gauge theory radiation manifestly exhibits color-kinematic duality.

Given Eqs.~(\ref{eq:snlo}),~(\ref{eq:yma}), the relation between the bi-adjoint and gauge theory observables is summarized  in terms of the formal substitution rule 
\begin{equation}
\label{eq:sdc}
{\tilde C}_i^{\tilde a} \mapsto N^\mu_i(k).
\end{equation}
Taking this one step further, one maps the remaining color structure in Eq.~(\ref{eq:yma}) according to the same substitution rule to obtain an object
\begin{equation}
{\cal A}^{\mu \nu}_n(k) = \kappa^n \sum_{ij} N^\mu_i(k) P^{ij}(k) N_j^\nu(k),
\end{equation}
which automatically obeys $k_\mu {\cal A}_n^{\mu\nu}= k_\nu {\cal A}_n^{\mu\nu}=0$ due to the explicit color-kinematic duality. By direct calculation~\cite{Shen:2018nlo}, this indeed reproduces the dilaton gravity amplitude ${\cal A}^{\mu\nu}$ at NLO.   One advantage of this formulation is that the numerators have manifest color-kinematic duality at any given order in perturbation theory, which automatically guaratees that the double copy amplitudes satisfy the Ward identities and hence define a consistent theory containing massless spin-2 particles.

As an example of this procedure, and for use in later sections, we recast the leading order results in terms of the factorization proposed by Shen. The independent color numerators in the LO bi-adjoint amplitude are
\begin{equation}
\label{eq:loc}
C^{a}=\left(\begin{array}{c}
(c_{\alpha}\cdot c_{\beta})c_{\alpha}^{a}\\{}
[c_{\alpha},c_{\beta}]^{a}
\end{array}\right),\ \tilde{C}^{\tilde{a}}=\left(\begin{array}{c}
(\tilde{c}_{\alpha}\cdot\tilde{c}_{\beta})\tilde{c}_{\alpha}^{\tilde{a}}\\{}
[\tilde{c}_{\alpha},\tilde{c}_{\beta}]^{\tilde{a}}
\end{array}\right),
\end{equation}
so from Eq.~(\ref{eq:bs}), we can read off the propagator matrix
\begin{equation}
P= -i\mu_{\alpha,\beta}(k)\left(\begin{array}{cc}
\ell_{\alpha}^{2}\frac{k\cdot\ell_{\beta}}{(k\cdot p_{\alpha})^{2}} & \frac{\ell_{\alpha}^{2}}{k\cdot p_{\alpha}}\\
\frac{\ell_{\alpha}^{2}}{k\cdot p_{\alpha}} & -1
\end{array}\right)\label{eq:prop},
\end{equation}
and the LO amplitude is ${\cal A}^{a\tilde a} = y^3 \sum_{\alpha,\beta} \int (C^a)^T P {\tilde C}^{\tilde a}$. Similarly, the gauge theory amplitude at LO can be decomposed as
\begin{equation}
{\cal A}^{a\mu}(k) = g^3 \sum_{\alpha,\beta} \int (C^a)^T P N^\mu,
\end{equation}
where the kinematic numerators are 
\begin{equation}
\label{eq:lon}
N^\mu=\left(\begin{array}{c}
(p_{\alpha}\cdot p_{\beta})p_{\alpha}^{\mu}\\
(k\cdot p_{\alpha})p_{\beta}^{\mu}-(k\cdot p_{\beta})p_{\alpha}^{\mu}+\frac{1}{2}(p_{\alpha}\cdot p_{\beta})\ell_{\alpha}^{\mu}-\frac{1}{2}(p_{\alpha}\cdot p_{\beta})\ell_{\beta}^{\mu}
\end{array}\right).
\end{equation}
   Notice from the form of the color factors $C^a$ and kinematic numerators $N^\mu$ in Eqs.~(\ref{eq:loc}),~(\ref{eq:lon}) that at this order in perturbation theory, the rules Eq.~(\ref{eq:rules}) originally proposed in~\cite{Goldberger:2016ymdc} are consistent with Eq.~(\ref{eq:sdc}). However, at NLO, Eq.~(\ref{eq:rules}) no longer yields consistent amplitude relations.

\section{The double copy for spinning sources and classical bosonic strings}\label{sec:spdc}
\subsection{Spinning sources and color-kinematic duality}

In this section we revisit the results of~\cite{Goldberger:2017cadc,Li:2018sgdc} on radiation from spinning point-like objects within the context of the results present in~\cite{Shen:2018nlo}.    We begin by noting that if the numerical value of the chromo-magnetic coupling constant is chosen precisely as in Eq.~(\ref{eq:dipole}), the Yang-Mills radiation field, constructed to linear order in spin, can be put into the form
\begin{equation}
\label{eq:yms1}
\mathcal{A}^{\mu,a}(k)=g^{3}\sum_{\alpha,\beta} \int ({C^a})^T  P {N_R}^\mu
\end{equation}
where the effects of spin are encoded in the new numerator structure $N_R = N + N_S$, with $N^\mu$ as in Eq.~(\ref{eq:lon}) and to linear order in spin,
\begin{equation}
{N_S}^\mu=i\left(\begin{array}{c}
(\ell_{\beta}\wedge p_{\alpha})_{\beta}p_{\alpha}^{\mu}-(\ell_{\beta}\wedge p_{\beta})_{\alpha}p_{\alpha}^{\mu}-(p_{\alpha}\cdot p_{\beta})(S_{\alpha}\wedge k)^{\mu}+(k\cdot p_{\alpha})(S_{\alpha}\wedge p_{\beta})^{\mu}\\
(k\cdot p_{\beta})(S_{\alpha}\wedge\ell_{\alpha})^{\mu}+(\ell_{\alpha}\wedge\ell_{\beta})_{\beta}p_{\alpha}^{\mu}-\frac{1}{2}(\ell_{\alpha}\wedge p_{\beta})_{\alpha}(\ell_{\beta}-\ell_{\alpha})^{\mu}-(\alpha\longleftrightarrow\beta).
\end{array}\right).\label{eq:spkin}
\end{equation}
In this equation, we have introduced the shorthand notation  $(S_{\alpha}\wedge p)^{\mu} = S_{\alpha}^{\mu\nu}p_{\nu},$ $(k\wedge p)_{\alpha} =k_{\mu}(S_{\alpha}\wedge p)^{\mu}$.  Remarkably, the propagator matrix is identical to Eq. (\ref{eq:prop}) found in the absence of spin, which is another way to understand why the choice of gyromagnetic ratio in Eq.~(\ref{eq:dipole}) is the natural one from the point of view of the double copy.    

Once we have the amplitude in the form of Eq.~(\ref{eq:yms1}) we can apply color-to-kinematics substitution rules to generate a gravitational amplitude.    First of all, the substitution $C^a\mapsto N^\mu$ in Eq.~(\ref{eq:yms1}) gives an amplitude
\begin{equation}
\mathcal{A}^{\mu\nu}(k)=\kappa^{3}\sum_{\alpha,\beta} \int (N^\nu)^T  P {N_R}^\mu+\mathcal{O}(S^{2}),\label{eq:splo},
\end{equation}
which is consistent with the results of~\cite{Goldberger:2017cadc,Li:2018sgdc} for $(\phi,g_{\mu\nu},B_{\mu\nu})$ radiation in the bulk theory defined by Eq.~(\ref{eq:sgrav}), sourced by spinning point particles that couple as in Eq.~(\ref{eq:pcp}).    However, the proposal in~\cite{Shen:2018nlo} allows a generalization of this mapping to a wider class of spinning extended objects. To see this, note that as in BCJ duality, the numerator structure in the color-to-kinematic replacements need not correspond to the same $N^\mu$ which we read off the gauge theory result. Rather, we can use the mapping $C^a\mapsto N^\mu_L$, where in general the consistent numerators  ${N_L}^\mu$ (with $k_\mu {N_L}^\mu=0$), do not coincide with ${N_R}^\mu$, but are instead taken from a different gauge theory radiation amplitude. The gravitational radiation field in this more general situation is given by 
\begin{equation}
\mathcal{A}^{\mu\nu}(k)=\kappa^{3}\sum_{\alpha,\beta} \int ({N_L}^\nu)^T  P {N_R}^\mu,
\end{equation}
which also corresponds to a consistent solutions, $k_\mu {\cal A}^{\mu\nu} = k_\nu {\cal A}^{\mu\nu}=0$.

In particular, formally we can take $N^\mu_R$ from a theory of point color charges carrying spin variables labeled $S^{\mu\nu}_R$ and $N^\mu_L$ from a system of particles with an independent set of spin degrees of freedom $S_L^{\mu\nu}$ but identical otherwise (same color charges and momenta).  The results of~\cite{Goldberger:2017cadc,Li:2018sgdc} are recovered if we take $S^{\mu\nu}_L=0$ and $S^{\mu\nu}_R=S^{\mu\nu}$, while taking $S^{\mu\nu}_L=S^{\mu\nu}$ and $S^{\mu\nu}_R=0$ flips the order of $\epsilon$ and $\tilde{\epsilon}$ leading to a sign change in the axion channel. This corresponds to coupling the spin to a different connection
\begin{equation}
S_{pp}\supset  {1\over 2} \int dx^\rho  {C^-}_{\mu\nu\rho} {\tilde S}^{\mu\nu},
\end{equation}
where ${C^-}^\rho_{(\mu\nu)} = {\tilde \Gamma}^\rho{}_{\mu\nu}$  and torsion $T^\rho{}_{\mu\nu} = {C^-}^\rho{}_{[\mu\nu]}= - {\tilde H}^\rho{}_{\mu\nu}$.     On the other hand, the choice
\begin{equation}
S_L^{\mu\nu} = S_R^{\mu\nu} = {1\over 2} S^{\mu\nu},\label{eq:unori}
\end{equation}
yields an amplitude ${\cal A}^{\mu\nu}(k) = {\cal A}^{\nu\mu}(k)$ corresponding to spinning particles with vanishing axion couplings.

\subsection{Classical bosonic strings}

As it turns out, the more general situation with independent spins $S_L^{\mu\nu}$ and $S_R^{\mu\nu}$ coupled to gravitational fields via a term  
\begin{equation}
S_{pp}\supset -\frac{1}{4}\int dx^{\lambda}(S_{L}^{\mu\nu}-S_{R}^{\mu\nu})H_{\mu\nu\lambda}e^{-2\phi}.\label{eq:spaxicp}
\end{equation}
also has a sensible physical interpretation.  In fact, the form of this coupling together with Eq.~(\ref{eq:dipole}) suggests a mapping between a classical open string coupled to a gauge field and a classical closed string interacting with massless fields $(\phi,g_{\mu\nu},B_{\mu\nu})$, at least to leading order in the multipole expansion and in the classical regime.

To provide evidence for this claim, we consider first an open string which is in a ``semi-classical'' configuration (i.e. it is in a state where at least some of the oscillator variables $\alpha^\mu_n$ are highly occupied and the string invariant mass $M_s$ is large in string units). If this object is placed in an external field $A_\mu$ whose typical time and distance scales are large compared to the string length\footnote{Our conventions are those of~\cite{Green:1986gsw}, $S={1\over 2\pi\ell^2}\int d^2 \sigma\sqrt{h} h^{ab}\partial_a X^\mu \partial_b X_\mu$.   We work in conformal gauge, $h_{ab}=\eta_{ab}$.}  $\ell,$ we can describe it systematically as a point source carrying gauge-invariant interactions with the gauge field.   We focus on the interactions linear in $A_\mu$ and consider for simplicity the case of $U(1)$ gauge symmetry. In the full theory of the extended string, the gauge field couples to Chan-Paton charges $q_\sigma$ localized at the string endpoints $\sigma=0,\pi$ ($g_o$ is the open string coupling)
\begin{equation}
\label{eq:cp}
S_{int} = g_o  \sum_{\sigma=0,\pi} q_\sigma \int dX^\mu(\tau,\sigma) A_\mu (X(\tau,\sigma)).
\end{equation}    

It is useful to work in the Polyakov gauge in which the string equations of motion in the absence of external fields are simply $(\partial_\tau^2 - \partial_\sigma^2) X^\mu =0$ subject to Virasoro constraints $\partial_\tau X \cdot \partial_\sigma X =0$, and $(\partial_\tau X)^2 = - (\partial_\sigma X)^2$.   Then to linear order in external fields, the point particle limit can be read off by inserting into Eq.~(\ref{eq:cp}) the solution to the classical string equations of motion,
\begin{equation}
X^\mu(\tau,\sigma) = x^\mu + \ell^2 p^\mu \tau + Z^\mu(\tau,\sigma),
\end{equation}
with oscillator contribution $Z^\mu= i\ell \sum_{n\neq 0} {\alpha^\mu_n\over n} e^{-in\tau} \cos n\sigma$, and expanding about the center-of-mass coordinate $x^\mu(\tau)= x^\mu + \ell^2 p^\mu \tau$.    In particular, the $U(1)$ current induced by the moving string is the ``vertex operator''
\begin{equation}
J^\mu(k)= \int d^d x e^{ik\cdot x} J^\mu(x) = g_o \sum_{\sigma=0,\pi} q_\sigma \int d\tau \partial_\tau X^\mu(\sigma,\tau) e^{i k\cdot X(\sigma,\tau)}, 
\end{equation}
 and the multipole limit corresponds to the formal expansion in powers of $k^\mu\rightarrow 0$ and $Z^\mu$.    Choosing coordinates with $x^\mu=0$, the leading order term in this expansion is  
 \begin{equation}
J^\mu(k\rightarrow 0) \simeq g_o  \sum_\sigma q_\sigma \int d\tau e^{i \ell^2 k\cdot p\tau} \ell^2 p^\mu+\mathcal{O}(Z^1,k\cdot Z^1) = g_s Q (2\pi) \delta(k\cdot p) p^\mu ,
 \end{equation}
 which is of course the current of a static point particle whose $U(1)$ charge is the sum $Q=\sum_\sigma q_\sigma$.   
 
 At the next order in the expansion, the string motion generates an electric dipole moment of the form
 \begin{equation}
\left. J^\mu(k\rightarrow 0)\right|_E \simeq -i g_o  \sum_\sigma q_\sigma \int d\tau e^{i \ell^2 k\cdot p\tau} \left(k\cdot p Z^\mu - (k\cdot Z) p^\mu\right)+\mathcal{O}(Z^2,k\cdot Z^2)
 \end{equation}
after integration by parts.    Inserting the mode expansion for $Z^\mu$, this is a sum of delta functions $\delta(\ell^2 k\cdot p -n)$ with $n\neq 0$ and therefore vanishes at frequencies $\omega\equiv k\cdot v\ll M_s^{-1}\ell^{-2}$.   So there is no permanent (time independent) electric dipole moment, as expected from time-reversal symmetry.   At the same order in the multipole expansion but next order in powers of $Z^\mu$, we also have a term
\begin{align}
\left. J^\mu(k\rightarrow 0)\right|_B &\simeq g_o\sum_\sigma q_\sigma \int d\tau e^{i \ell^2 k\cdot p\tau} ik\cdot Z \partial_\tau Z^\mu+\mathcal{O}(k\cdot Z^2)\\&\simeq -{i\over 2} g_o k_\nu \sum_\sigma q_\sigma \int d\tau e^{i \ell^2 k\cdot p\tau} \left(Z^\mu \partial_\tau Z^\nu - Z^\nu \partial_\tau Z^\mu\right)+\mathcal{O}(k\cdot Z^2)
\end{align}
Here we have discarded a term which, upon integration by parts, acquires a factor of $(k\cdot p)$ and is suppressed in the static limit $\omega\ll M_s^{-1}\ell^{-2}$.    Given that 
\begin{equation}
\left. Z^\mu {\dot Z}^\nu\right|_{\sigma=0,\pi} = -i \ell^2 \sum_{n\neq 0} {1\over n}  \alpha^\mu_n \alpha^\nu_{-n} + \mbox{time dependent terms},
\end{equation}
the part of the moment that remains after rapid oscillations of order the string scale are averaged out is proportional to the angular momentum of the string about the center of mass $x^\mu=0$,
\begin{equation}
\left. J^\mu(k\rightarrow 0)\right|_B  = i g_o Q (2\pi) \delta(k\cdot p) k_\nu S^{\mu\nu}
\end{equation}
with intrinsic spin $S^{\mu\nu}\gg \hbar$ given in terms of the oscillator modes by~\cite{Green:1986gsw}
\begin{equation}
S^{\mu\nu} =  \int_0^\pi d\sigma \left(Z^\mu \partial_\tau Z^\nu - Z^\nu \partial_\tau Z^\mu\right) = - i \sum_{n=1}^\infty {1\over n} \left(\alpha^\mu_{-n} \alpha^\nu_{n}- \alpha_{-n}^{\nu}\alpha_{n}^{\mu}\right).
\end{equation}

To summarize, we have found the current induced by the open string in the long-wavelength limit
\begin{equation}
J^\mu(k) = g_o Q (2\pi)\delta(k\cdot p) \left[1 +  i k_\nu S^{\mu\nu}\right]+\cdots.
\end{equation}
This is precisely the form one finds for a pointlike startic particle with current $J^\mu(x) = \delta S_{pp}/\delta A_\mu(x)$ and 
\begin{equation}
S_{pp} = g_o Q \int dx^\mu A_\mu(x) + {1\over 2} g_o Q  \int d\tau S^{\mu\nu} F_{\mu\nu}(x).
\end{equation}
In particular, in $d=4$ dimensions, if we make the identification $g_o=g$, the spin-dependent terms corresponds to a magnetic moment interaction with gyromagnetic ratio equal to Dirac's value $g_D=2$, consistent with earlier classical~\cite{Ademollo:1974te} and and quantum mechanical~\cite{Ferrara:1992gfac} string theory results.     The non-abelian generalization of this calculation should proceed along the same lines, with a Wilson line inserted between the string endpoints in order to ensure gauge invariance.     The result in that case is then the chromomagnetic interaction in Eq.~(\ref{eq:dipole}) with the same numerical value suggested by the consistency of the classical double copy.

Similarly, we can perform the multipole expansion of a closed oriented string coupled to a long-wavelength axion field.    The relevant interaction term in the worldsheet action is now
\begin{equation}
S\supset  -{1\over 2\pi \ell^2} \int d^2 \sigma \epsilon^{ab} B_{\mu\nu}(X) \partial_a X^\mu \partial_b X^\nu.
\end{equation}
In this case, the mode expansions subject to periodic closed string boundary conditions  in $\sigma\mapsto\sigma+\pi$ are given by
\begin{equation}
Z^\mu(\tau,\sigma)= Z^\mu_L(\tau+\sigma) +Z^\mu_R(\tau-\sigma)\equiv i\ell\sum_{n\neq0}\frac{1}{2n}e^{2in\sigma}(e^{-2in\tau}\alpha_{n}^{\mu}-e^{2in\tau}\tilde{\alpha}_{-n}^{\mu}),
\end{equation}
representing the contribution of left- and right-moving oscillators. In momentum space, the current that couples to the axion corresponds to the vertex operator
\begin{equation}
J^{\mu\nu}(k) = -g_{c}\int d\tau \int_{0}^{\pi}d\sigma e^{ik\cdot X}\left(\partial_\tau X^{\mu}\partial_\sigma X^{\nu}-\partial_\sigma X^{\mu}\partial_\tau X^{\nu}\right),
\end{equation}
where the closed string coupling is $g_c\equiv{1\over 2\pi \ell^2}$. Again, expanding in $Z^\mu$ and $k\cdot Z$, we find that the leading term is formally linear in $Z^\mu$ but vanishes due to the periodic boundary condition. Thus, the string has zero ``axion charge'' (monopole moment). At the next order, integrating over $\sigma$, and discarding the contribution from the non-zero modes, proportional to $\delta(\ell^2 k\cdot p -n)$ with $n\neq0$, we find that
\begin{equation}
\label{eq:pc}
J^{\mu\nu}(k)=-\frac{i}{2}(2\pi)\delta(k\cdot p)k_{\lambda}(S_L^{\mu\nu}-S_R^{\mu\nu}).
\end{equation}
where the spins in the two sectors are
\begin{align}
S_{L}^{\mu\nu}=-i\sum_{n=1}^{\infty}\frac{1}{n}(\alpha_{-n}^{\mu}\alpha_{n}^{\nu}-\alpha_{-n}^{\nu}\alpha_{n}^{\mu}),\\
S_{R}^{\mu\nu}=-i\sum_{n=1}^{\infty}\frac{1}{n}(\tilde{\alpha}_{-n}^{\mu}\tilde{\alpha}_{n}^{\nu}-\tilde{\alpha}_{-n}^{\nu}\tilde{\alpha}_{n}^{\mu}).
\end{align}
Eq.~(\ref{eq:pc}) matches the axion current induced by a point object which couples to $B_{\mu\nu}$ as
\begin{equation}
S_{pp}\supset {1\over 4} \int dx^\rho H_{\rho\mu\nu}(x) (S^{\mu\nu}_L-S^{\mu\nu}_R),
\end{equation}
which agrees precisely with the prediction of the double copy.    This differs from Eq. (\ref{eq:spaxicp}) by its dependence on the dilaton, but recalling that the present analysis is done in the ``string frame'', it is understood that for $\phi\neq0$ the field strength should be identified with $\tilde{H}$ instead.    

Finally, we note that under worldsheet parity, we have $\Omega\ :\ S_{R}\mapsto S_{L}$ and $S_{pp}\mapsto -S_{pp}$ as expect for oriented string configurations.    From this point of view, the parity even correspondence  $S_L^{\mu\nu}=S_R^{\mu\nu}={1\over 2}S^{\mu\nu}$ is analogous to the case of unoriented strings, which do not couple to $B_{\mu\nu}$.

\section{Dynamical polarizability and its double copy\label{sec:hdim}}

\begin{figure}
	\begin{centering}
		\includegraphics[scale=0.5]{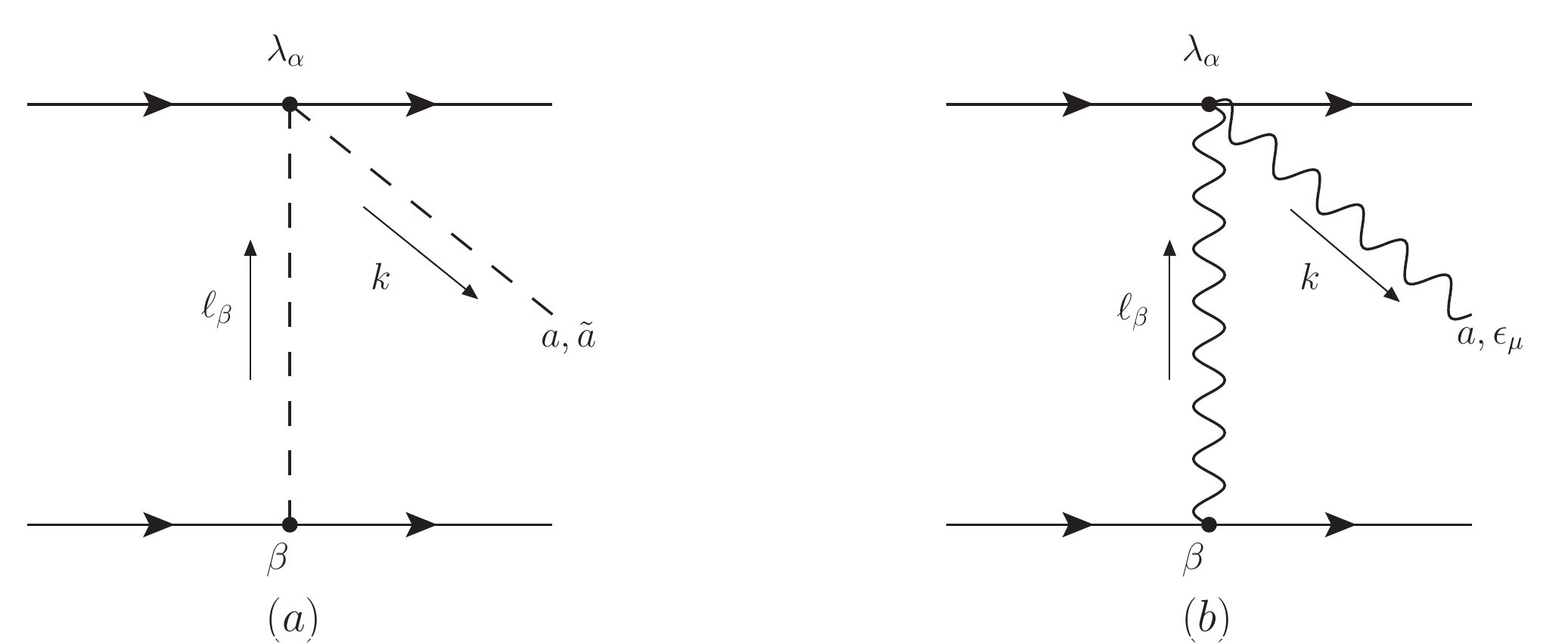}
		\par\end{centering}
	\caption{The leading order diagrams for bi-adjoint and Yang-Mills radiation
		amplitudes induced by finite-size worldline operators, (a) for bi-adjoint scalar and
		(b) for Yang-Mills.\label{fig:bsymlorad-1} }
	
\end{figure}

The discussion above can also be generalized to objects which have finite extent $\sim {\cal R}$.    If probed by long-wavelength external fields, $\lambda\gg {\cal R}$ such objects can still be described by a worldline theory.  In addition to the minimal couplings in Eqs.~(\ref{eq:bapp})~(\ref{eq:ympp})~(\ref{eq:gpp}) the theory also contains higher dimensional operators involving more derivatives and/or powers of the external fields.   By dimensional analysis, the Wilson coefficients of the non-minimal operators scale as powers of the size of the extended object, giving rise to effects suppressed by powers of ${\cal R}/\lambda\ll1$.   By including all such operators that are allowed by the symmetries of the theory, we systematically account for all possible finite-size effects\footnote{This neglects the possibility of absorption, for instance by black hole horizons, which requires the existence of new nearly gapless worldline localized degrees of freedom.    See~\cite{Goldberger:2005disseff}.}, order by order in the power counting.    We now show how such finite-size effects are constrained by demanding consistency of the double copy.

At leading (quadratic in fields) order, the finite-size operators determine linear response, namely the multipole moments induced on the finite-size object by an external field configuration.   In the bi-adjoint theory, the linear response operators consistent with $G\times {\tilde G}$ symmetry arise at zeroth order in spacetime derivatives, and consist of the four terms
\begin{equation}
\label{eq:bfs}
S={1\over 2} y^2 \sum_i \lambda_i \int d\tau {\cal O}^{BS}_{(i)},
\end{equation}
where 
\begin{eqnarray}
\begin{array}{cccc}
{\cal O}^{BS}_{(1)} = \phi^{a{\tilde a}} \phi^{a{\tilde a}},  & {\cal O}^{BS}_{(2)} = (c\cdot \phi)^{\tilde a} (c\cdot \phi)^{\tilde a}, &
 {\cal O}^{BS}_{(3)} =   (c\cdot \phi)^{\tilde a} (c\cdot \phi)^{\tilde a},  &    {\cal O}^{BS}_{(4)} =(c\cdot \phi \cdot {\tilde c})^2,
\end{array}
\end{eqnarray}
and $\lambda^\alpha_i$ are set of dimensionless coupling constants.  For clarity we have suppressed the particle label $\alpha$ on the coupling constants and action.  We have also introduced the shorthand notation $(c\cdot \phi)^{\tilde a}= c_a \phi^{a{\tilde a}},$ $(\phi\cdot {\tilde c})^{ a}=\phi^{a{\tilde a}} {\tilde c}_{\tilde a}$, and $c\cdot \phi \cdot {\tilde c} = c_a {\tilde c}_{\tilde a} \phi^{a\tilde a}$. We normalize these operators with prefactor $y^2$ and impose the coupling constant mapping rules $y\mapsto g\mapsto \kappa$ under the double copy.

It is straightforward to compute the contribution to long-distance radiation from the terms in Eq.~(\ref{eq:bfs}).   We work to linear order in the parameters $\lambda_i$ and consider a scattering event where the field generated by a second point source deforms the extended object.   The time-dependence of the induced moments then sources scalar radiation.   Diagrammatically, the situation is depicted in Fig.~\ref{fig:bsymlorad-1}(a). To linear order in the finite-size couplings, the amplitude is given by
\begin{equation}
\mathcal{A}^{a\tilde{a}}=y^{3}\sum_{\alpha,\beta}\int {\tilde C}_a^T \Lambda C_a,\label{eq:bsfs}
\end{equation}
where we have defined the color numerators as
\begin{equation}
C^{a}=\left(\begin{array}{c}
c_{\beta}^{a}\\
(c_{\alpha}\cdot c_{\beta})c_{\alpha}^{a}
\end{array}\right),\ \tilde{C}^{\tilde{a}}=\left(\begin{array}{c}
\tilde{c}_{\beta}^{\tilde{a}}\\
(\tilde{c}_{\alpha}\cdot\tilde{c}_{\beta})\tilde{c}_{\alpha}^{\tilde{a}}
\end{array}\right),\label{eq:fscol}
\end{equation}
as well as the $2\times 2$ matrix $\Lambda_\alpha$ of Wilson coefficients and a propagator prefactor for particle $\alpha$
\begin{equation}
\Lambda =
\mu_{\alpha,\beta}(k)\ell_{\alpha}^{2}\left(\begin{array}{cc}
\lambda^\alpha_1 & \lambda^\alpha_3\\
\lambda^\alpha_2 & \lambda^\alpha_4
\end{array}\right).
\end{equation}

Likewise, in gauge theory,  the linear response of a (spinless) extended color sources in the long-wavelength limit can be described by including the operators
\begin{equation}
S = {1\over 4} g^2 \sum_ i \lambda_i \int d\tau {\cal O}^{YM}_{(i)},
\end{equation}
where now
\begin{equation}
\begin{array}{cccc}
{\cal O}^{YM}_{(1)}= F_{\mu\nu}^{a}F_{a}^{\mu\nu}, & {\cal O}^{YM}_{(2)}= (c\cdot F)_{\mu\nu} (c\cdot F)^{\mu\nu}, & {\cal O}^{YM}_{(3)} = {\dot x}^\mu {\dot x}^\nu  F^a_{\mu\sigma}   {F^a_{\nu}}^{\sigma}, &  {\cal O}^{YM}_{(4)} =  {\dot x}^\mu {\dot x}^\nu (c\cdot F)_{\mu\sigma}  (c\cdot F)_{\nu}^{\ \sigma}.
\end{array}
\end{equation}
These operators induce contributions to radiation in scattering as shown in Fig.~\ref{fig:bsymlorad-1}(b).  By explicit calculation, we find that the result exhibits a factorization property which parallels that found in the bi-adjoint case,
\begin{equation}
\mathcal{A}^{a\mu}=g^{3}\sum_{\alpha,\beta}\int{\tilde C}_a^T \Lambda_\alpha N^\mu.\label{eq:ymfs}
\end{equation}
where $C_i^a$, $i=1,2$ is the set of color factors given in Eq.~(\ref{eq:fscol}) while the kinematic numerators are defined to be
\begin{equation}
N^\mu=\left(\begin{array}{c}
(k\cdot p_{\beta})\ell_{\beta}^{\mu}-(k\cdot\ell_{\beta})p_{\beta}^{\mu}\\
\frac{1}{2}(p_{\alpha}\cdot p_{\beta})[(k\cdot p_{\alpha})\ell_{\beta}^{\mu}-(k\cdot\ell_{\beta})p_{\alpha}^{\mu}]+\frac{1}{2}(k\cdot p_{\alpha})[(k\cdot p_{\beta})p_{\alpha}^{\mu}-(k\cdot p_{\alpha})p_{\beta}^{\mu}]
\end{array}\right).\label{eq:fsslkn}
\end{equation}

In order to account for the full finite-size effects in gravity (including all the axionic operators), it is also necessary to consider the radiation induced by the interactions of the spin of the probe particle $\beta$ with the extended object $\alpha$.  Working to linear order in spin\footnote{We ignore finite-size operators built out of spin at this order in gauge or finite size couplings. It is easy to see that there is no kinematic numerator with a dual representation to linear order in spin. However, we might still need include such terms at higher orders in perturbation theory.}, the amplitude is represented by diagrams of the same topology as in Fig.~\ref{fig:bsymlorad-1}(b) where the off-shell gluon is emitted from the chromomagnetic coupling of particle $\beta$. The result is then of the same form as Eq.~(\ref{eq:ymfs}), but the kinematic factor is shifted by a term linear in $S^{\mu\nu}_\beta$, $N^\mu\rightarrow N^\mu +  {N_S}^\mu,$ where
\begin{equation}
{N_S}^\mu=i\left(\begin{array}{c}
(k\cdot\ell_{\beta})(S_{\beta}\wedge\ell_{\beta})^{\mu}-(\ell_{\alpha}\wedge\ell_{\beta})_{\beta}\ell_{\beta}^{\mu}\\
\frac{1}{2}(\ell_{\beta}\wedge p_{\alpha})_{\beta}[(k\cdot p_{\alpha})\ell_{\beta}^{\mu}-(k\cdot\ell_{\beta})p_{\alpha}^{\mu}]+\frac{1}{2}(k\cdot p_{\alpha})[(k\cdot p_{\alpha})(S_{\beta}\wedge\ell_{\beta})^{\mu}-(\ell_{\alpha}\wedge\ell_{\beta})_{\beta}p_{\alpha}^{\mu}]
\end{array}\right).\label{eq:fsspkn}
\end{equation}

The respective results in Eq.~(\ref{eq:bsfs}) and Eq.~(\ref{eq:ymfs}) suggest a set of single copy mapping rules between finite-size objects in bi-adjoint scalar and Yang-Mills theory.    Namely, making the replacement ${\tilde C}^a\mapsto N^\mu$ maps the finite-size amplitude ${\cal A}^{a\tilde a}$ of bi-adjoint theory to the radiation field ${\cal A}^{a\mu}$ in gauge theory\footnote{Notice that this mapping takes operators with no derivatives on $\phi^{a\tilde a}$ to operators involving gradients of $A^\mu_a$ in Yang-Mills. Including an operator of the form, e.g.,  $y^2\int d\tau (\partial_\mu \phi^{a\tilde a})^2$ in scalar theory yields a radiation amplitude $\mathcal{A}^{a\tilde a}= y^3 \int \mu_{\alpha,\beta}\ell_{\alpha}^{2}(k\cdot\ell_{\beta}) {\tilde C}^{\tilde a}_1 C_1^a$ whose propagator structure $\ell_{\alpha}^{2}(k\cdot\ell_{\beta})$ does not match with any of the terms in Eq.~(\ref{eq:ymfs})   Rather, it corresponds to a four-derivative operator $\int d\tau D_{\sigma}F_{\mu\nu}^{a}D^{\sigma}F_{a}^{\mu\nu}$ which yields an amplitude of the form $g^3\int \mu_{\alpha,\beta}\ell_{\alpha}^{2}(k\cdot\ell_{\beta}) C_1^a N^\mu_1$ consistent with the color-kinematics substitution ${\tilde C}^{\tilde a}\mapsto N^\mu$.}. We note that the Wilson coefficients in Eq.~(\ref{eq:bsfs}) and Eq.~(\ref{eq:ymfs}) are in principle different. However, this mapping relation gives a direct correspondence between the two sets $\lambda_i\mapsto\lambda^{YM}_i\equiv\lambda_i$. Given this, it is then natural to take a further step ${\cal A}^{a\mu}\mapsto {\cal A}^{\mu\nu}$ while demanding $\lambda_i$ unchanged, where 
\begin{equation}
\mathcal{A}^{\mu\nu}=\kappa^{3}\sum_{\alpha,\beta}\int ({N_L}^\nu)^T \Lambda {N_R}^\mu,
\end{equation}
with ${N_{L,R}}^\mu={N_{S_{L,R}}}^\mu$.  Because $k_\mu \mathcal{A}^{\mu\nu} = k_\nu \mathcal{A}^{\mu\nu}=0$ this defines a consistent radiation field in a theory of finite-size sources coupled to massless fields $(\phi,g_{\mu\nu},B_{\mu\nu})$.

Given the structure of the single copy ${\cal A}^{a\tilde{a}}\mapsto {\cal A}^{a\mu}$, which takes finite-size operators in bi-adjoint theory with no derivatives to two-derivative operators in gauge theory, we expect that the double copy amplitude ${\cal A}^{\mu\nu}$ encodes finite-size effects corresponding to a total of four derivatives acting on the fields $(\phi,g_{\mu\nu},B_{\mu\nu})$.  To determine the precise form of the finite-size response encoded in ${\cal A}^{\mu\nu}$, we therefore start with the most general set of four-derivative diffeomorphism invariant worldline operators that are quadratic in these fields.  Since we are not considering spin-dependent finite-size operators in the gauge theory, we also limit ourselves to spin-independent gravitational higher dimensional operators.  At the four-derivative level, the complete set of terms allowed by diffeomorphism invariance is
\begin{equation}
S=\sum_i\tilde{\lambda}_i\int d\tau {\cal \tilde{O}}_{(i)},\label{eq:fsgrav}
\end{equation}
where now we have twelve Wilson coefficients $\tilde{\lambda}_i$ corresponding to ten positive definite operators
\begin{align}
{\cal \tilde{O}}_{(G_1)}={1\over 4} (R_{\mu\nu\rho\sigma})^2,  {\cal \tilde{O}}_{(G_2)}&={1\over 4} (R_{\mu\nu\rho\sigma}\dot{x}^{\sigma})^2, {\cal \tilde{O}}_{(G_3)}= {1\over 4} (R_{\mu\nu\rho\sigma}\dot{x}^{\nu}\dot{x}^{\sigma})^2,\label{eq:gterms}\\
{\cal \tilde{O}}_{(D_1)}={1\over 2} (\nabla_{\mu}\nabla_{\nu}\phi)^2,  {\cal \tilde{O}}_{(D_2)}&= {1\over 2}(\dot{x}^\mu \nabla_\mu \nabla_\nu\phi)^2,  {\cal \tilde{O}}_{(D_3)} = {1\over 2}(\dot{x}^\mu \dot{x}^\nu\nabla_\mu \nabla_\nu\phi)^2,\label{eq:dterms}\\
{\cal \tilde{O}}_{(A_1)}= {1\over 6}(\nabla_{\sigma}H_{\mu\nu\rho})^2, {\cal \tilde{O}}_{(A_2)}={1\over 4}(\nabla_{\sigma}H_{\mu\nu\rho}&\dot{x}^{\rho})^2,   {\cal \tilde{O}}_{(A_3)}={1\over 6} (\nabla_{\sigma}H_{\mu\nu\rho}\dot{x}^{\sigma})^2,  {\cal \tilde{O}}_{(A_4)}={1\over 4} (\nabla_{\sigma}H_{\mu\nu\rho}\dot{x}_{}^{\sigma}\dot{x}_{}^{\rho})^2,\label{eq:aterms}
\end{align}
and to two terms that mix the graviton with the dilaton or axion
\begin{align}
{\cal \tilde{O}}_{(GD)}&=(\dot{x}^\rho \dot{x}^\sigma R_{\mu\rho\nu\sigma})\nabla^{\mu}\nabla^{\nu}\phi,\label{eq:gdterms}\\
{\cal \tilde{O}}_{(GA)}&=(\dot{x}^{\sigma} R_{\mu\nu\rho\sigma})(\dot{x}_{\lambda}\nabla^{\mu}H^{\nu\rho\lambda}).\label{eq:gaterms}
\end{align}
Note that we have omitted operators involving the Ricci tensor $R_{\mu\nu}$, since by field redefinitions these are equivalent on-shell to terms constructed out of derivatives of $\phi$ and $B_{\mu\nu}$.

\begin{figure}
	\begin{centering}
		\includegraphics[scale=0.5]{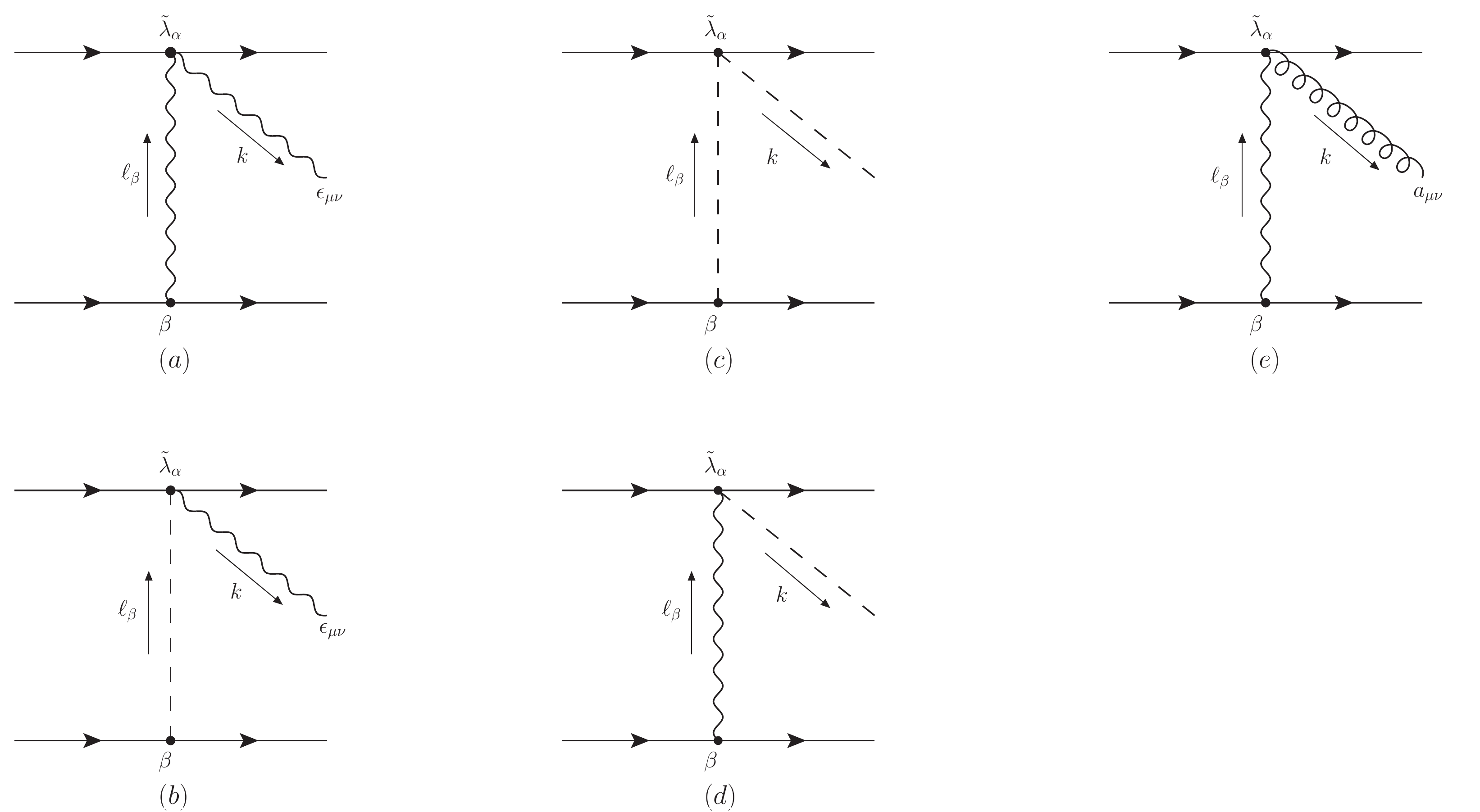}
		\par\end{centering}
	\caption{The leading order radiation for spinless sources with finite-size corrections. The dotted lines represent dilatons, the coiled lines are axions, and the thin wavy lines mean gravitons. The first column
		contributes to graviton channel, the second are responsible for dilaton
		channel, and the third gives the radiation in axion channel.\label{fig:glorad}}
\end{figure}

\begin{figure}
	\begin{centering}
		\includegraphics[scale=0.5]{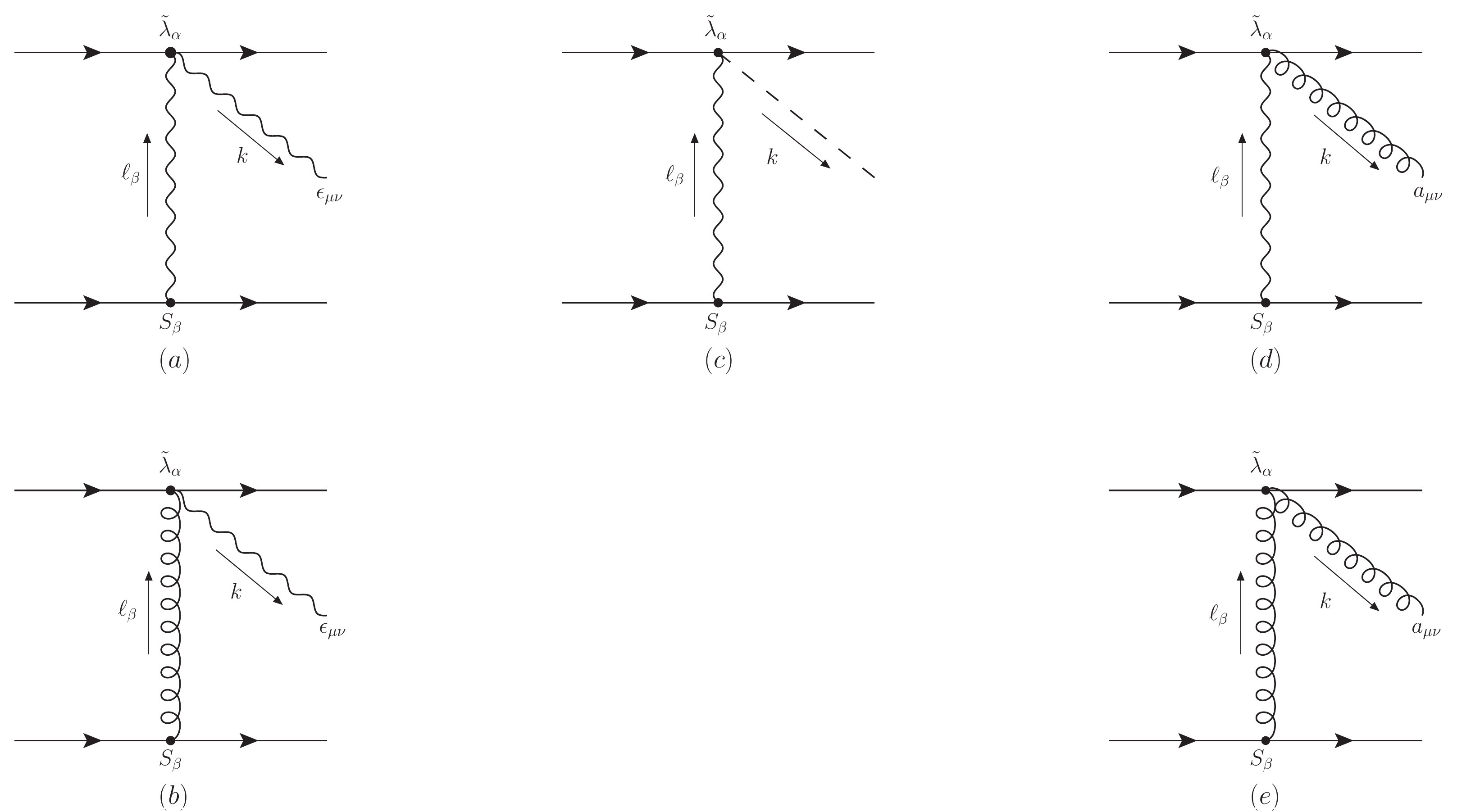}
		\par\end{centering}
	\caption{Diagrammatics for leading order finite-size contributions from spinning sources at linear order in spin.
		The same symbols are used to represent the respective fields.\label{fig:fssp}}
	\end{figure}

The amplitudes corresponding to radiation from the induced multipoles at zeroth and first orders in spin are calculated from the diagrams in Figs.~\ref{fig:glorad},~\ref{fig:fssp} respectively. It turns out that the individual amplitudes corresponding to each of the operators in Eq.~(\ref{eq:fsgrav}) do not factorize in the way that would be expected from color-kinematics.    However, by taking linear combinations of operators,  it is possible to construct amplitudes where the only kinematic numerators that arise coincide with those that appear in the gauge theory (in Eqs.~(\ref{eq:fsslkn}),~(\ref{eq:fsspkn}).    For this choice of operators coefficients, the amplitude in the gravity theory agrees with the prediction of the double copy given by Eq.~(\ref{eq:ymfs}).    

  As an explicit example, we consider the case with positive axion-spin coupling, $S^{\mu\nu}_L=0$ and $S^{\mu\nu}_R=S^{\mu\nu}$. The explicit calculations are reported in Appendix~\ref{app:sp}.    The result is that the gravitational Wilson coefficients are related to the finite-size coupling on the gauge theory side by the relations
  \begin{align}
&\lambda_{1}^\alpha	=2\tilde{\lambda}_{G_1}^\alpha=\frac{\tilde{\lambda}_{D_1}^\alpha}{d-2}=4\tilde{\lambda}_{A_1}^\alpha,\label{eq:lam1}\\
\lambda_{2}^\alpha	=2\tilde{\lambda}_{G_2}^\alpha=-2\tilde{\lambda}&_{GD}^\alpha=2\tilde{\lambda}_{D_1}^\alpha=\frac{2\tilde{\lambda}_{D_2}^\alpha}{d-4}=4\tilde{\lambda}_{GA}^\alpha=4\tilde{\lambda}_{A_2}^\alpha=8\tilde{\lambda}_{A_3}^\alpha,\label{eq:lam2}\\
\lambda_{3}^\alpha	=2\tilde{\lambda}_{G_2}^\alpha=-2\tilde{\lambda}&_{GD}^\alpha=2\tilde{\lambda}_{D_1}^\alpha=\frac{2\tilde{\lambda}_{D_2}^\alpha}{d-4}=-4\tilde{\lambda}_{GA}^\alpha=4\tilde{\lambda}_{A_2}^\alpha=8\tilde{\lambda}_{A_3}^\alpha,\label{eq:lam3}\\
\lambda_{4}^\alpha	=2\tilde{\lambda}_{G_3}^\alpha=&-4\tilde{\lambda}_{GD}^\alpha=4\tilde{\lambda}_{D_1}^\alpha=-2\tilde{\lambda}_{D_2}^\alpha=\frac{4\tilde{\lambda}_{D_4}^\alpha}{d-2}=4\tilde{\lambda}_{A_4}^\alpha.\label{eq:lam4}
\end{align}
We note that although the full set of operators in the string gravity background are not independent, there are still more free coefficients than the number of purely gravitational operators. Thus it should be possible to characterize the full gravitational tidal response at the linear level, provided one could project out the fields $\phi,B_{\mu\nu}$ in a systematic way that does not introduce new constraints among the gravitational tidal operators.

It is also interesting to note that, at least for some of these relations, there is a geometrical pattern.    In terms of the non-minimal connection $C^+_{\mu\nu\rho}$ connection defined in Eq.~(\ref{eq:pcp}),  the above relations imply that the independent operators in the gravitational double can be expressed as
\begin{equation}
S=\frac{1}{8}\sum_i\lambda_i\int d\tau {\cal O}^{SG}_{(i)},
\end{equation}
where
\begin{align}
\mathcal{O}^{SG}_{(1)}&=(\tilde{R}^+_{\mu\nu\rho\sigma})^2,\\
\mathcal{O}^{SG}_{(2)}=(\tilde{R}^+_{\mu\nu\rho\sigma}\dot{x}^\sigma)^2+&{\cal \tilde{O}}_{(GA)}+\frac{1}{4}\tilde{O}_{(A_2)}+\frac{1}{12}\tilde{O}_{(A_3)},\\
\mathcal{O}^{SG}_{(3)}=(\tilde{R}^+_{\mu\nu\rho\sigma}\dot{x}^\sigma)^2-3&{\cal \tilde{O}}_{(GA)}+\frac{1}{4}\tilde{O}_{(A_2)}+\frac{1}{12}\tilde{O}_{(A_3)},\\
\mathcal{O}^{SG}_{(4)}&=(\tilde{R}^+_{\mu\nu\rho\sigma}\dot{x}^\rho\dot{x}^\sigma)^2.
\end{align}
However, while the dilaton dependence has been completely absorbed into the curvature associated with the non-minimal connection $C^+_{\mu\nu\rho}$ , it does not seem possible to modify the torsion in order to simplify the axion dependent terms.

\section{Discussions and conclusions\label{sec:disc}}

In this paper, we have generalized the classical double copy formulation of~\cite{Shen:2018nlo}, based on manifest color-kinematic duality, beyond the strict point particle limit.    In particular, we have shown that this yields a consistent map between radiation amplitudes in the case of spinning particles and for extended objects, in the limit where their response to long-wavelength probe fields can be encoded in a non-minimal worldline effective theory.   Of course, this paper only includes calculations at the lowest order where the color structures are extremely simple. To justify the color-kinematic duality beyond the point particle limit, we would have to consider higher-order contributions, where more complicated color factors and (non-diagonal) propagator matrices will arise.

One of our findings in this paper is that the double copy substitution rules applied to spinning gauge theory sources allow for consistent objects, on the gravity side, carrying pairs of spins $S_L^{\mu\nu}$, $S_R^{\mu\nu}$ on their worldlines.  We have found that such objects have a possible interpretation\footnote{Another interpretation can be found in ref.~\cite{Bautista:2019evw}, which identified the point sources on the gravity side with massive states arising in the compactification of ${\cal N}=4$ supergravity.} as a long-wavelength limit of a highly massive (classical) closed string.   Evidence for this claim includes the fact that on both the gauge theory and gravity side of the duality, the sources must have couplings to massless fields whose numerical values precisely match those predicted by string theory.   The picture this suggests is of classical open strings on the Yang-Mills side getting mapped via color-kinematics duality to closed strings interacting with the $(\phi,g_{\mu\nu},B_{\mu\nu})$ fields that naturally arise in the double copy.    According to this interpretation, identifying the two spins as $S_L^{\mu\nu}=S_R^{\mu\nu}={1\over 2} S^{\mu\nu}$ then corresponds to the case of unoriented strings, with $B_{\mu\nu}$ decoupled.   

It would be interesting to pursue this connection to strings, not unexpected given the close relation between BCJ duality and the earlier work of KLT, in more detail.   Doing so might help explain the pattern of constraints we found on the finite-size Wilson coefficients in Eqs.~(\ref{eq:lam1})-(\ref{eq:lam4}).   Apart from the partial simplification that occurs by working with the Weyl-rescaled metric, it is natural to guess that the constraints also arise from the internal structure of the string sources.       Similarly, the constraint on operators in the case, $S_L^{\mu\nu}=S_R^{\mu\nu}$, including the decoupling of the axion to all orders, might have some connection with a limit of the unoriented string.

Phenomenologically, the inclusion of higher-dimensional operators into the double copy map enables one to study more realistic source objects with finite size (e.g. with non-zero tidal Love numbers).   However, making contact with phenomenology still requires a systematic method for obtaining pure Einstein gravity.   In this paper, we have found a way to remove the axion, at least at leading order in perturbation theory.   Some possibilities for removing the dilaton have been proposed, for instance  by introducing ghost fields~\cite{Luna:2017lnow} or by direct construction in terms of scattering amplitude relations~\cite{Bautista:2019sca}.  It would be useful to test if any of these approaches can be extended to incorporate finite-size corrections as well.    We leave these considerations for future work.

\section{Acknowledgement}

This work was supported in part by the US Department of Energy under Grant No. DE­SC0017660.

\begin{appendix}

\section{Radiation from higher dimensional operators}\label{app:fsamp}

We collect in the appendix the explicit gravity calculations leading to the results quoted in Sec.~\ref{sec:hdim}.   First we consider radiation induced by graviton or dilaton exchange only.   Then, as depicted in Fig. \ref{fig:glorad}(a), the quadratic graviton operators Eq.~(\ref{eq:gterms}) lead to graviton radiation amplitudes
	\begin{align}
	\mathcal{A}^{\mu\nu}_{G_1}= & 2\kappa^3\sum_{\alpha,\beta}\int_{\ell_{\alpha,\beta}}\mu_{\alpha,\beta}(k)\ell_{\alpha}^{2}\bigg[(k\cdot p_{\beta})\ell_{\beta}^{\mu}-(k\cdot\ell_{\beta})p_{\beta}^{\mu}\bigg]\bigg[(k\cdot p_{\beta})\ell_{\beta}^{\nu}-(k\cdot\ell_{\beta})p_{\beta}^{\nu}\bigg],\label{eq:g1}\\
	\mathcal{A}^{\mu\nu}_{G_2}= & \kappa^3\sum_{\alpha,\beta}\int_{\ell_{\alpha,\beta}}\mu_{\alpha,\beta}(k)\ell_{\alpha}^{2}\Bigg\{\bigg[(p_{\alpha}\cdot p_{\beta})[(k\cdot p_{\alpha})\ell_{\beta}^{\mu}-(k\cdot\ell_{\beta})p_{\alpha}^{\mu}]+(k\cdot p_{\alpha})[(k\cdot p_{\beta})p_{\alpha}^{\mu}-(k\cdot p_{\alpha})p_{\beta}^{\mu}]\bigg]\nonumber \\
	& \times\bigg[(k\cdot p_{\beta})\ell_{\beta}^{\nu}-(k\cdot\ell_{\beta})p_{\beta}^{\nu}\bigg]-\frac{1}{d-2}\bigg[(k\cdot p_{\alpha})\ell_{\beta}^{\mu}-(k\cdot\ell_{\beta})p_{\alpha}^{\mu}\bigg]\bigg[(k\cdot p_{\alpha})\ell_{\beta}^{\nu}-(k\cdot\ell_{\beta})p_{\alpha}^{\nu}\bigg]\Bigg\},\label{eq:g2}\\
	\mathcal{A}^{\mu\nu}_{G_3}= & \frac{\kappa^3}{2}\sum_{\alpha,\beta}\int_{\ell_{\alpha,\beta}}\mu_{\alpha,\beta}(k)\ell_{\alpha}^{2}\Bigg\{\bigg[(p_{\alpha}\cdot p_{\beta})[(k\cdot p_{\alpha})\ell_{\beta}^{\mu}-(k\cdot\ell_{\beta})p_{\alpha}^{\mu}]+(k\cdot p_{\alpha})[(k\cdot p_{\beta})p_{\alpha}^{\mu}-(k\cdot p_{\alpha})p_{\beta}^{\mu}]\bigg]\nonumber \\
	& \times\bigg[(p_{\alpha}\cdot p_{\beta})[(k\cdot p_{\alpha})\ell_{\beta}^{\nu}-(k\cdot\ell_{\beta})p_{\alpha}^{\nu}]+(k\cdot p_{\alpha})[(k\cdot p_{\beta})p_{\alpha}^{\nu}-(k\cdot p_{\alpha})p_{\beta}^{\nu}]\bigg]\nonumber \\
	& -\frac{1}{d-2}\bigg[(k\cdot p_{\alpha})\ell_{\beta}^{\mu}-(k\cdot\ell_{\beta})p_{\alpha}^{\mu}\bigg]\bigg[(k\cdot p_{\alpha})\ell_{\beta}^{\nu}-(k\cdot\ell_{\beta})p_{\alpha}^{\nu}\bigg]\Bigg\}.\label{eq:g3}
	\end{align}
In addition, the graviton-dilaton operator Eq.~(\ref{eq:gdterms}) gives rise to the diagram Fig. \ref{fig:glorad}(b) that corresponds to graviton radiation of the form
\begin{equation}
\mathcal{A}^{\mu\nu}_{G_D}=-\frac{\kappa^3}{d-2}\sum_{\alpha,\beta}\int_{\ell_{\alpha,\beta}}\mu_{\alpha,\beta}(k)\ell_{\alpha}^{2}\bigg[(k\cdot p_{\alpha})\ell_{\beta}^{\mu}-(k\cdot\ell_{\beta})p_{\alpha}^{\mu}\bigg]\bigg[(k\cdot p_{\alpha})\ell_{\beta}^{\nu}-(k\cdot\ell_{\beta})p_{\alpha}^{\nu}\bigg].
\end{equation}
We observe that the combinations $\mathcal{A}_{G_2}-\mathcal{A}_{G_D}$ and $\mathcal{A}_{G_3}-\frac{1}{2}\mathcal{A}_{G_D}$ yield results independent of $d$.

In the dilaton channel, the quadratic dilaton operators Eq.~(\ref{eq:dterms}) allow diagrams of the form shown in Fig.~\ref{fig:glorad}(c), resulting in radiation amplitudes
	\begin{align}
	\mathcal{A}_{D_1}= & \frac{\kappa^3}{(d-2)^{3/2}}\sum_{\alpha,\beta}\int_{\ell_{\alpha,\beta}}\mu_{\alpha,\beta}(k)\ell_{\alpha}^{2}(k\cdot\ell_{\beta})^{2},\label{eq:d1}\\
	\mathcal{A}_{D_2}= & \frac{\kappa^3}{(d-2)^{3/2}}\sum_{\alpha,\beta}\int_{\ell_{\alpha,\beta}}\mu_{\alpha,\beta}(k)\ell_{\alpha}^{2}(k\cdot\ell_{\beta})(k\cdot p_{\alpha})^{2},\label{eq:d2}\\
	\mathcal{A}_{D_3}= & \frac{\kappa^3}{(d-2)^{3/2}}\sum_{\alpha,\beta}\int_{\ell_{\alpha,\beta}}\mu_{\alpha,\beta}(k)\ell_{\alpha}^{2}(k\cdot p_{\alpha})^{4}.\label{eq:d3}
	\end{align}
In addition, Fig.~\ref{fig:glorad}(d) with an insertion of the graviton-dilaton operator Eq.~(\ref{eq:gdterms}) also leads to radiation in the scalar channel,
\begin{equation}
\mathcal{A}_{D_G}=\frac{\kappa^3}{(d-2)^{1/2}}\sum_{\alpha,\beta}\int_{\ell_{\alpha,\beta}}\mu_{\alpha,\beta}(k)\ell_{\alpha}^{2}\Bigg[\big((p_{\alpha}\cdot p_{\beta})(k\cdot\ell_{\beta})-(k\cdot p_{\alpha})(k\cdot p_{\beta})\big)^{2}-\frac{(k\cdot\ell_{\beta})^{2}}{d-2}+\frac{2(k\cdot\ell_{\beta})(k\cdot p_{\alpha})}{d-2}\Bigg].
\end{equation}
Finally, diagram Fig. \ref{fig:glorad}(e) with a single insertion of the graviton-axion operator Eq.~(\ref{eq:gaterms}) yields the axion radiation
amplitude
\begin{equation}
\mathcal{A}^{\mu\nu}_{A_G}=2\kappa^3\int_{\ell_{\alpha,\beta}}\mu_{\alpha,\beta}(k)\ell_{\alpha}^{2}\bigg[(p_{\alpha}\cdot p_{\beta})[(k\cdot p_{\alpha})\ell_{\beta}^{\nu}-(k\cdot\ell_{\beta})p_{\alpha}^{\nu}]+(k\cdot p_{\alpha})[(k\cdot p_{\beta})p_{\alpha}^{\nu}-(k\cdot p_{\alpha})p_{\beta}^{\nu}]\bigg]\bigg[(k\cdot p_{\beta})\ell_{\beta}^{\mu}-(k\cdot\ell_{\beta})p_{\beta}^{\mu}\bigg].
\end{equation}

We note that in matching the double copy to the dilaton channel, we have omitted contact terms with no propagator factors.    Such terms yield integrals that are proportional to
to
\begin{align}
& \int\frac{d^{d}\ell_{\alpha}}{(2\pi)^{d}}\frac{d^{d}\ell_{\beta}}{(2\pi)^{d}}\bigg[(2\pi)\delta(\ell_{\alpha}\cdot p_{\alpha})e^{i\ell_{\alpha}\cdot b_{\alpha}}\bigg]\bigg[(2\pi)\delta(\ell_{\alpha}\cdot p_{\alpha})e^{i\ell_{\beta}\cdot b_{\beta}}\bigg](2\pi)^{d}\delta^{d}(\ell_{\alpha}+\ell_{\beta}-k)\nonumber \\
= & \int d\tau_{\alpha}d\tau_{\beta}e^{ik\cdot x_{\beta}^{(0)}}\int\frac{d^{d}\ell}{(2\pi)^{d}}e^{i\ell\cdot x_{\alpha}^{(0)}}e^{-i\ell\cdot x_{\beta}^{(0)}}=\int d\tau_{\alpha}d\tau_{\beta}e^{ik\cdot x_{\beta}^{(0)}}\delta^{d}(x_{\alpha}^{(0)}-x_{\beta}^{(0)}),
\end{align}
where the free particle paths are $x_{\alpha}^{(0)}=b_{\alpha}+v_{\alpha}\tau_{\alpha}$.    Because we consider classical scattering at non-zero impact parameter $b_{\alpha\beta}=b_\alpha-b_\beta\neq 0$, such terms are identically zero.

\subsection{Spin-dependent terms}\label{app:sp}
The spinning point sources now support internal graviton and axion exchange, corresponding to the diagrams in Fig. \ref{fig:fssp}(a).  These yield graviton emission amplitudes	
	\begin{align}
	\mathcal{A}^{\mu\nu}_{G_1}= & 2i\kappa^3\sum_{\alpha,\beta}\int_{\ell_{\alpha,\beta}}\mu_{\alpha,\beta}(k)\ell_{\alpha}^{2}\bigg[(k\cdot p_{\beta})\ell_{\beta}^{\nu}-(k\cdot\ell_{\beta})p_{\beta}^{\nu}\bigg]\bigg[(k\cdot\ell_{\beta})(S_{\beta}\wedge\ell_{\beta})^{\mu}-(\ell_{\alpha}\wedge\ell_{\beta})_{\beta}\ell_{\beta}^{\mu}\bigg],\label{eq:gs1}\\
	\mathcal{A}^{\mu\nu}_{G_2}= & \frac{i\kappa^3}{2}\sum_{\alpha,\beta}\int_{\ell_{\alpha,\beta}}\mu_{\alpha,\beta}(k)\ell_{\alpha}^{2}\Bigg\{\bigg[(k\cdot p_{\beta})\ell_{\beta}^{\nu}-(k\cdot\ell_{\beta})p_{\beta}^{\nu}\bigg]\nonumber \\
	\times & \bigg[(\ell_{\beta}\wedge p_{\alpha})_{\beta}[(k\cdot p_{\alpha})\ell_{\beta}^{\mu}-(k\cdot\ell_{\beta})p_{\alpha}^{\mu}]+(k\cdot p_{\alpha})[(k\cdot p_{\alpha})(S_{\beta}\wedge\ell_{\beta})^{\mu}-(\ell_{\alpha}\wedge\ell_{\beta})_{\beta}p_{\alpha}^{\mu}]\bigg]\nonumber \\
	+\bigg[(p_{\alpha} & \cdot p_{\beta})[(k\cdot p_{\alpha})\ell_{\beta}^{\nu}-(k\cdot\ell_{\beta})p_{\alpha}^{\nu}]+(k\cdot p_{\alpha})[(k\cdot p_{\beta})p_{\alpha}^{\nu}-(k\cdot p_{\alpha})p_{\beta}^{\nu}]\bigg]\bigg[(k\cdot\ell_{\beta})(S_{\beta}\wedge\ell_{\beta})^{\mu}-(\ell_{\alpha}\wedge\ell_{\beta})_{\beta}\ell_{\beta}^{\mu}\bigg]\Bigg\},\label{eq:gs2}\\
	\mathcal{A}^{\mu\nu}_{G_3}= & \frac{i\kappa^3}{2}\sum_{\alpha,\beta}\int_{\ell_{\alpha,\beta}}\mu_{\alpha,\beta}(k)\ell_{\alpha}^{2}\bigg[(p_{\alpha}\cdot p_{\beta})[(k\cdot p_{\alpha})\ell_{\beta}^{\nu}-(k\cdot\ell_{\beta})p_{\alpha}^{\nu}]+(k\cdot p_{\alpha})[(k\cdot p_{\beta})p_{\alpha}^{\nu}-(k\cdot p_{\alpha})p_{\beta}^{\nu}]\bigg]\nonumber \\
	& \times\bigg[(\ell_{\beta}\wedge p_{\alpha})_{\beta}[(k\cdot p_{\alpha})\ell_{\beta}^{\mu}-(k\cdot\ell_{\beta})p_{\alpha}^{\mu}]+(k\cdot p_{\alpha})[(k\cdot p_{\alpha})(S_{\beta}\wedge\ell_{\beta})^{\mu}-(\ell_{\alpha}\wedge\ell_{\beta})_{\beta}p_{\alpha}^{\mu}]\bigg].\label{eq:gs3}
	\end{align}
At linear order in spin, we can also have graviton radiation mediated by axion exhange, as in Fig.~\ref{fig:fssp}(b)
\begin{align}
\mathcal{A}^{\mu\nu}_{G_A}= & i\kappa^3\sum_{\alpha,\beta}\int_{\ell_{\alpha,\beta}}\mu_{\alpha,\beta}(k)\ell_{\alpha}^{2}\Bigg\{-\bigg[(k\cdot p_{\beta})\ell_{\beta}^{\nu}-(k\cdot\ell_{\beta})p_{\beta}^{\nu}\bigg]\nonumber \\
\times & \bigg[(\ell_{\beta}\wedge p_{\alpha})_{\beta}[(k\cdot p_{\alpha})\ell_{\beta}^{\mu}-(k\cdot\ell_{\beta})p_{\alpha}^{\mu}]+(k\cdot p_{\alpha})[(k\cdot p_{\alpha})(S_{\beta}\wedge\ell_{\beta})^{\mu}-(\ell_{\alpha}\wedge\ell_{\beta})_{\beta}p_{\alpha}^{\mu}]\bigg]\nonumber \\
+\bigg[(p_{\alpha} & \cdot p_{\beta})[(k\cdot p_{\alpha})\ell_{\beta}^{\nu}-(k\cdot\ell_{\beta})p_{\alpha}^{\nu}]+(k\cdot p_{\alpha})[(k\cdot p_{\beta})p_{\alpha}^{\nu}-(k\cdot p_{\alpha})p_{\beta}^{\nu}]\bigg]\bigg[(k\cdot\ell_{\beta})(S_{\beta}\wedge\ell_{\beta})^{\mu}-(\ell_{\alpha}\wedge\ell_{\beta})_{\beta}\ell_{\beta}^{\mu}\bigg]\Bigg\}.
\end{align}
There is also dilaton radiation,  from the diagram in Fig.~\ref{fig:fssp}(c) with one insertion of the graviton-dilaton mixing operator in Eq.~(\ref{eq:gdterms})
\begin{equation}
\mathcal{A}_{D_G}=\frac{i\kappa^3}{(d-2)^{1/2}}\sum_{\alpha,\beta}\int_{\ell_{\alpha,\beta}}\mu_{\alpha,\beta}(k)\ell_{\alpha}^{2}\bigg[(\ell_{\beta}\wedge p_{\alpha})_{\beta}(k\cdot\ell_{\beta})+(\ell_{\alpha}\wedge\ell_{\beta})_{\beta}(k\cdot p_{\alpha})\bigg]\bigg[(k\cdot p_{\beta})(k\cdot p_{\alpha})-(k\cdot\ell_{\beta})(p_{\alpha}\cdot p_{\beta})\bigg].
\end{equation}

Finally, there is spin-dependent axion radiation, involving insertions of the purely axionic operators in Eq.~(\ref{eq:aterms}) in the diagram of Fig. \ref{fig:fssp}(e).   Two of these amplitudes readily factorize into products of the kinematic factors appearing in gauge theory
\begin{align}
	\mathcal{A}^{\mu\nu}_{A_1}= & 4i\kappa^3\sum_{\alpha,\beta}\int_{\ell_{\alpha,\beta}}\mu_{\alpha,\beta}(k)\ell_{\alpha}^{2}\bigg[(k\cdot p_{\beta})\ell_{\beta}^{\nu}-(k\cdot\ell_{\beta})p_{\beta}^{\nu}\bigg]\bigg[(k\cdot\ell_{\beta})(S_{\beta}\wedge\ell_{\beta})^{\mu}-(\ell_{\alpha}\wedge\ell_{\beta})_{\beta}\ell_{\beta}^{\mu}\bigg],\\
	\mathcal{A}^{\mu\nu}_{A_4}= & 2i\kappa^3\sum_{\alpha,\beta}\int_{\ell_{\alpha,\beta}}\mu_{\alpha,\beta}(k)\ell_{\alpha}^{2}\bigg[(p_{\alpha}\cdot p_{\beta})[(k\cdot p_{\alpha})\ell_{\beta}^{\nu}-(k\cdot\ell_{\beta})p_{\alpha}^{\nu}]+(k\cdot p_{\alpha})[(k\cdot p_{\beta})p_{\alpha}^{\nu}-(k\cdot p_{\alpha})p_{\beta}^{\nu}]\bigg]\nonumber \\
	& \times\bigg[(\ell_{\beta}\wedge p_{\alpha})_{\beta}[(k\cdot p_{\alpha})\ell_{\beta}^{\mu}-(k\cdot\ell_{\beta})p_{\alpha}^{\mu}]+(k\cdot p_{\alpha})[(k\cdot p_{\alpha})(S_{\beta}\wedge\ell_{\beta})^{\mu}-(\ell_{\alpha}\wedge\ell_{\beta})_{\beta}p_{\alpha}^{\mu}]\bigg].
	\end{align}
The remaining two, $\mathcal{A}_{A_2}$ and $\mathcal{A}_{A_3}$, do not factorize into Yang-Mills kinematic factors.    However, if we also include $\mathcal{A}_{A_G}$ (Fig. \ref{fig:fssp}(d)) obtained from inserting the graviton-axion operator Eq.~(\ref{eq:gaterms}), we find that the linear combinations
\begin{align}
\mathcal{A}^{\mu\nu}_{A_G}+\mathcal{A}^{\mu\nu}_{A_2}+\frac{1}{2}\mathcal{A}^{\mu\nu}_{A_3}= & 2i\kappa^3\sum_{\alpha,\beta}\int_{\ell_{\alpha,\beta}}\mu_{\alpha,\beta}(k)\ell_{\alpha}^{2}\bigg[(k\cdot\ell_{\beta})(S_{\beta}\wedge\ell_{\beta})^{\mu}-(\ell_{\alpha}\wedge\ell_{\beta})_{\beta}\ell_{\beta}^{\mu}\bigg]\nonumber \\
& \times\bigg[(p_{\alpha}\cdot p_{\beta})[(k\cdot p_{\alpha})\ell_{\beta}^{\nu}-(k\cdot\ell_{\beta})p_{\alpha}^{\nu}]+(k\cdot p_{\alpha})[(k\cdot p_{\beta})p_{\alpha}^{\nu}-(k\cdot p_{\alpha})p_{\beta}^{\nu}]\bigg],\\
\mathcal{A}^{\mu\nu}_{A_G}-\mathcal{A}^{\mu\nu}_{A_2}-\frac{1}{2}\mathcal{A}^{\mu\nu}_{A_3}= & -2i\kappa^3\sum_{\alpha,\beta}\int_{\ell_{\alpha,\beta}}\mu_{\alpha,\beta}(k)\ell_{\alpha}^{2}\bigg[(k\cdot p_{\beta})\ell_{\beta}^{\nu}-(k\cdot\ell_{\beta})p_{\beta}^{\nu}\bigg]\nonumber \\
\times\bigg[(\ell_{\beta} & \wedge p_{\alpha})_{\beta}[(k\cdot p_{\alpha})\ell_{\beta}^{\mu}-(k\cdot\ell_{\beta})p_{\alpha}^{\mu}]+(k\cdot p_{\alpha})[(k\cdot p_{\alpha})(S_{\beta}\wedge\ell_{\beta})^{\mu}-(\ell_{\alpha}\wedge\ell_{\beta})_{\beta}p_{\alpha}^{\mu}]\bigg],
\end{align}
indeed factorize.

\end{appendix}


\begin{references}
\bibitem{Bern:2008bcj1}
Z.~Bern, J.~J.~M.~Carrasco and H.~Johansson,
Phys.\ Rev.\ D {\bf 78}, 085011 (2008)
[arXiv:0805.3993 [hep-ph]].

\bibitem{Bern:2010bdhk}
Z.~Bern, T.~Dennen, Y.~t.~Huang and M.~Kiermaier,
Phys.\ Rev.\ D {\bf 82}, 065003 (2010)
[arXiv:1004.0693 [hep-th]].

\bibitem{Kawai:1986klt}
H.~Kawai, D.~C.~Lewellen and S.~H.~H.~Tye,
Nucl.\ Phys.\ B {\bf 269}, 1 (1986).

\bibitem{proof}
N.~E.~J.~Bjerrum-Bohr, P.~H.~Damgaard, T.~Sondergaard and P.~Vanhove,
JHEP {\bf 1101}, 001 (2011)
[arXiv:1010.3933 [hep-th]];
C.~R.~Mafra, O.~Schlotterer and S.~Stieberger,
JHEP {\bf 1107}, 092 (2011)
[arXiv:1104.5224 [hep-th]].

\bibitem{Bern:2010bcj2}
Z.~Bern, J.~J.~M.~Carrasco and H.~Johansson,
Phys.\ Rev.\ Lett.\  {\bf 105}, 061602 (2010)
[arXiv:1004.0476 [hep-th]].

\bibitem{fiveloops}
Z.~Bern, J.~J.~Carrasco, W.~M.~Chen, A.~Edison, H.~Johansson, J.~Parra-Martinez, R.~Roiban and M.~Zeng,
Phys.\ Rev.\ D {\bf 98}, no. 8, 086021 (2018)
[arXiv:1804.09311 [hep-th]];
Z.~Bern, J.~J.~M.~Carrasco, W.~M.~Chen, H.~Johansson, R.~Roiban and M.~Zeng,
Phys.\ Rev.\ D {\bf 96}, no. 12, 126012 (2017)
[arXiv:1708.06807 [hep-th]].

\bibitem{Bern:2019prr} 
Z.~Bern, J.~J.~Carrasco, M.~Chiodaroli, H.~Johansson and R.~Roiban,
arXiv:1909.01358 [hep-th].

\bibitem{Monteiro:2014mow}
R.~Monteiro, D.~O'Connell and C.~D.~White,
JHEP {\bf 1412}, 056 (2014)
[arXiv:1410.0239 [hep-th]].

\bibitem{Luna:2015lmow}
A.~Luna, R.~Monteiro, D.~O'Connell and C.~D.~White,
Phys.\ Lett.\ B {\bf 750}, 272 (2015)
[arXiv:1507.01869 [hep-th]].


\bibitem{Luna:2016hge} 
A.~Luna, R.~Monteiro, I.~Nicholson, A.~Ochirov, D.~O'Connell, N.~Westerberg and C.~D.~White,
JHEP {\bf 1704}, 069 (2017)
[arXiv:1611.07508 [hep-th]];
K.~Kim, K.~Lee, R.~Monteiro, I.~Nicholson and D.~Peinador Veiga,
arXiv:1912.02177 [hep-th].




\bibitem{ksdc}
A.~K.~Ridgway and M.~B.~Wise,
Phys.\ Rev.\ D {\bf 94}, no. 4, 044023 (2016)
[arXiv:1512.02243 [hep-th]];
A.~Luna, R.~Monteiro, I.~Nicholson, D.~O'Connell and C.~D.~White,
JHEP {\bf 1606}, 023 (2016)
[arXiv:1603.05737 [hep-th]];
N.~Bahjat-Abbas, A.~Luna and C.~D.~White,
JHEP {\bf 1712}, 004 (2017)
[arXiv:1710.01953 [hep-th]].

\bibitem{ksdc2}
M.~Carrillo-Gonz\'alez, R.~Penco and M.~Trodden,
JHEP {\bf 1804}, 028 (2018)
[arXiv:1711.01296 [hep-th]];
M.~Gurses and B.~Tekin,
Phys.\ Rev.\ D {\bf 98}, no. 12, 126017 (2018)
[arXiv:1810.03411 [gr-qc]].



\bibitem{wave}
T.~Adamo, E.~Casali, L.~Mason and S.~Nekovar,
Class.\ Quant.\ Grav.\  {\bf 35}, no. 1, 015004 (2018)
[arXiv:1706.08925 [hep-th]];
A.~Ilderton,
Phys.\ Lett.\ B {\bf 782}, 22 (2018)
[arXiv:1804.07290 [gr-qc]];
T.~Adamo, E.~Casali, L.~Mason and S.~Nekovar,
JHEP {\bf 1902}, 198 (2019)
[arXiv:1810.05115 [hep-th]].

\bibitem{biadj}
C. D. White, C.~D.~White,
Phys.\ Lett.\ B {\bf 763}, 365 (2016)
[arXiv:1606.04724 [hep-th]];
P.~J.~De Smet and C.~D.~White,
Phys.\ Lett.\ B {\bf 775}, 163 (2017)
[arXiv:1708.01103 [hep-th]];
N.~Bahjat-Abbas, R.~Stark-Much\~ao and C.~D.~White,
Phys.\ Lett.\ B {\bf 788}, 274 (2019)
[arXiv:1810.08118 [hep-th]].

\bibitem{other}
K.~Lee,
JHEP {\bf 1810}, 027 (2018)
[arXiv:1807.08443 [hep-th]];
D.~S.~Berman, E.~Chac\'on, A.~Luna and C.~D.~White,
JHEP {\bf 1901}, 107 (2019)
[arXiv:1809.04063 [hep-th]];
M.~Carrillo Gonz\'alez, B.~Melcher, K.~Ratliff, S.~Watson and C.~D.~White,
JHEP {\bf 1907}, 167 (2019)
[arXiv:1904.11001 [hep-th]];

\bibitem{White:2017rev}
C.~D.~White,
Contemp.\ Phys.\  {\bf 59}, 109 (2018)
[arXiv:1708.07056 [hep-th]].

\bibitem{ligovirgo}
B.~P.~Abbott {\it et al.} [LIGO Scientific and Virgo Collaborations],
Phys.\ Rev.\ Lett.\  {\bf 116}, no. 6, 061102 (2016)
[arXiv:1602.03837 [gr-qc]];
B.~P.~Abbott {\it et al.} [LIGO Scientific and Virgo Collaborations],
Phys.\ Rev.\ Lett.\  {\bf 119}, no. 16, 161101 (2017)
[arXiv:1710.05832 [gr-qc]].

\bibitem{CRS}
C.~Cheung, I.~Z.~Rothstein and M.~P.~Solon,
Phys.\ Rev.\ Lett.\  {\bf 121}, no. 25, 251101 (2018)
[arXiv:1808.02489 [hep-th]].

\bibitem{Neill:2013wsa} 
D.~Neill and I.~Z.~Rothstein,
Nucl.\ Phys.\ B {\bf 877}, 177 (2013)
[arXiv:1304.7263 [hep-th]].

\bibitem{Vaidya:2014kza} 
V.~Vaidya,
Phys.\ Rev.\ D {\bf 91}, no. 2, 024017 (2015)
[arXiv:1410.5348 [hep-th]].

\bibitem{claspot}
Z.~Bern, C.~Cheung, R.~Roiban, C.~H.~Shen, M.~P.~Solon and M.~Zeng,
Phys.\ Rev.\ Lett.\  {\bf 122}, no. 20, 201603 (2019)
[arXiv:1901.04424 [hep-th]];
Z.~Bern, C.~Cheung, R.~Roiban, C.~H.~Shen, M.~P.~Solon and M.~Zeng,
JHEP {\bf 1910}, 206 (2019)
[arXiv:1908.01493 [hep-th]].

\bibitem{scattering}
N.~E.~J.~Bjerrum-Bohr, P.~H.~Damgaard, G.~Festuccia, L.~Planté and P.~Vanhove,
Phys.\ Rev.\ Lett.\  {\bf 121}, no. 17, 171601 (2018)
[arXiv:1806.04920 [hep-th]];
A.~Cristofoli, N.~E.~J.~Bjerrum-Bohr, P.~H.~Damgaard and P.~Vanhove,
Phys.\ Rev.\ D {\bf 100}, no. 8, 084040 (2019)
[arXiv:1906.01579 [hep-th]];
N.~E.~J.~Bjerrum-Bohr, A.~Cristofoli, P.~H.~Damgaard and H.~Gomez,
arXiv:1908.09755 [hep-th].

\bibitem{Goldberger:2016ymdc}
W.~D.~Goldberger and A.~K.~Ridgway,
Phys.\ Rev.\ D {\bf 95}, no. 12, 125010 (2017)
[arXiv:1611.03493 [hep-th]].



\bibitem{Goldberger:2006gr}
W.~D.~Goldberger and I.~Z.~Rothstein,
Phys.\ Rev.\ D {\bf 73}, 104029 (2006)
[hep-th/0409156].



\bibitem{NRGRrev}
W.~D.~Goldberger,
hep-ph/0701129;
S.~Foffa and R.~Sturani,
Class.\ Quant.\ Grav.\  {\bf 31}, no. 4, 043001 (2014)
[arXiv:1309.3474 [gr-qc]];
R.~A.~Porto,
Phys.\ Rept.\  {\bf 633}, 1 (2016)
[arXiv:1601.04914 [hep-th]];
M.~Levi,
arXiv:1807.01699 [hep-th].

\bibitem{Goldberger:2017bsdc}
W.~D.~Goldberger, S.~G.~Prabhu and J.~O.~Thompson,
Phys.\ Rev.\ D {\bf 96}, no. 6, 065009 (2017)
[arXiv:1705.09263 [hep-th]].

\bibitem{biadjoint}
Z.~Bern, A.~De Freitas and H.~L.~Wong,
Phys.\ Rev.\ Lett.\  {\bf 84}, 3531 (2000)
[hep-th/9912033];
R.~Monteiro and D.~O'Connell,
JHEP {\bf 1107}, 007 (2011)
[arXiv:1105.2565 [hep-th]];
Y.~J.~Du, B.~Feng and C.~H.~Fu,
JHEP {\bf 1108}, 129 (2011)
[arXiv:1105.3503 [hep-th]];
N.~E.~J.~Bjerrum-Bohr, P.~H.~Damgaard, R.~Monteiro and D.~O'Connell,
JHEP {\bf 1206}, 061 (2012)
[arXiv:1203.0944 [hep-th]];
F.~Cachazo, S.~He and E.~Y.~Yuan,
JHEP {\bf 1407}, 033 (2014)
[arXiv:1309.0885 [hep-th]];
A.~Anastasiou, L.~Borsten, M.~J.~Duff, L.~J.~Hughes and 	S.~Nagy,
Phys.\ Rev.\ Lett.\  {\bf 113}, no. 23, 231606 (2014)
[arXiv:1408.4434 [hep-th]].



\bibitem{Goldberger:2017cadc}
W.~D.~Goldberger, J.~Li and S.~G.~Prabhu,
Phys.\ Rev.\ D {\bf 97}, no. 10, 105018 (2018)
[arXiv:1712.09250 [hep-th]].

\bibitem{Li:2018sgdc}
J.~Li and S.~G.~Prabhu,
Phys.\ Rev.\ D {\bf 97}, no. 10, 105019 (2018)
[arXiv:1803.02405 [hep-th]].


\bibitem{Goldberger:2017bs}
W.~D.~Goldberger and A.~K.~Ridgway,
Phys.\ Rev.\ D {\bf 97}, no. 8, 085019 (2018)
[arXiv:1711.09493 [hep-th]]


\bibitem{Plefka:2018psw}
J.~Plefka, J.~Steinhoff and W.~Wormsbecher,
Phys.\ Rev.\ D {\bf 99}, no. 2, 024021 (2019)
[arXiv:1807.09859 [hep-th]];
J.~Plefka, C.~Shi, J.~Steinhoff and T.~Wang,
Phys.\ Rev.\ D {\bf 100}, no. 8, 086006 (2019)
[arXiv:1906.05875 [hep-th]].



\bibitem{Luna:2017lnow}
A.~Luna, I.~Nicholson, D.~O'Connell and C.~D.~White,
JHEP {\bf 1803}, 044 (2018)
[arXiv:1711.03901 [hep-th]].

\bibitem{Athira:2019}
A.~PV and A.~Manu,
arXiv:1907.10021 [hep-th].







\bibitem{clasobs}
D.~A.~Kosower, B.~Maybee and D.~O'Connell,
JHEP {\bf 1902}, 137 (2019)
[arXiv:1811.10950 [hep-th]].

\bibitem{scatangle} N.~E.~J.~Bjerrum-Bohr, A.~Cristofoli and P.~H.~Damgaard,
arXiv:1910.09366 [hep-th].


\bibitem{spin1}
A.~Guevara, A.~Ochirov and J.~Vines,
JHEP {\bf 1909}, 056 (2019)
[arXiv:1812.06895 [hep-th]];
A.~Guevara, A.~Ochirov and J.~Vines,
Phys.\ Rev.\ D {\bf 100}, no. 10, 104024 (2019)
[arXiv:1906.10071 [hep-th]].

\bibitem{Bautista:2019sca}
Y.~F.~Bautista and A.~Guevara,
arXiv:1903.12419 [hep-th];

\bibitem{spin2} 
B.~Maybee, D.~O'Connell and J.~Vines,
arXiv:1906.09260 [hep-th];
N.~Arkani-Hamed, Y.~t.~Huang and D.~O'Connell,
arXiv:1906.10100 [hep-th].


\bibitem{Bautista:2019evw} 
Y.~F.~Bautista and A.~Guevara,
arXiv:1908.11349 [hep-th].

\bibitem{Kalin:2019rwq} 
G.~K\"{a}lin and R.~A.~Porto,
arXiv:1910.03008 [hep-th];
arXiv:1911.09130 [hep-th].


\bibitem{Shen:2018nlo}
C.~H.~Shen,
JHEP {\bf 1811}, 162 (2018)
[arXiv:1806.07388 [hep-th]].


\bibitem{Bern:2016bdn}
Z.~Bern, S.~Davies and J.~Nohle,
Phys.\ Rev.\ D {\bf 93}, no. 10, 105015 (2016)
[arXiv:1510.03448 [hep-th]].

\bibitem{Green:1986gsw}
M.~B.~Green, J.~H.~Schwarz and E.~Witten,
Cambridge, Uk: Univ. Pr. ( 1987) 469 P. ( Cambridge Monographs On Mathematical Physics)

\bibitem{Ademollo:1974te} 
M.~Ademollo {\it et al.},
Nuovo Cim.\ A {\bf 21}, 77 (1974).

\bibitem{Ferrara:1992gfac}
S.~Ferrara, M.~Porrati and V.~L.~Telegdi,
Phys.\ Rev.\ D {\bf 46}, 3529 (1992).




\bibitem{Goldberger:2005disseff}W.~D.~Goldberger and I.~Z.~Rothstein,
Phys.\ Rev.\ D {\bf 73}, 104030 (2006)
[hep-th/0511133].

	
	

\end{references}
\end{document}